\documentclass[12pt]{article}
\usepackage[numbers]{natbib}
\usepackage{amsmath}
\usepackage{graphicx}
\usepackage{enumerate}
\usepackage{url} 
\usepackage{amsfonts}
\usepackage{amssymb}
\usepackage{mathrsfs}
\usepackage{yfonts}
\usepackage{subfig}
\usepackage{enumitem}
\usepackage{multicol}
\usepackage{multirow}
\usepackage{comment}

\usepackage[toc,page]{appendix}
\usepackage{hyperref}
\usepackage{amsthm} 
\usepackage{color}
\usepackage{colortbl}
\usepackage{mathtools}
\usepackage{subfig}
\usepackage[export]{adjustbox}

\usepackage{float}

\usepackage[english]{babel}

\usepackage{graphicx}
\usepackage{caption}

\setcitestyle{authoryear, open={[},close={]}}

\newcommand{\blind}{1}

\begin{document}

\def\spacingset#1{\renewcommand{\baselinestretch}%
{#1}\small\normalsize} \spacingset{1}


\if1\blind
{
  \title{\bf Cross Validation for Correlated Data in Regression and Classification Models, with Applications to Deep Learning}
  
  \author{Oren Yuval\thanks{
    The authors gratefully acknowledge \textit{the Israeli Science Foundation grant 2180/20} and support from \textit{the Israel Council for Higher Education Data-Science Centers}. This research has received funding from the European Union’s Horizon 2020 Framework Program
for Research and Innovation under the Specific Grant Agreement No. 945539 (Human Brain Project SGA3).}\hspace{.2cm}\\
    Department of Statistics and Operations Research, Tel-Aviv University\\
    and \\
    Saharon Rosset \\
    Department of Statistics and Operations Research, Tel-Aviv University}
  \maketitle
} \fi

\bigskip
\begin{abstract}
We present a methodology for model evaluation and selection where the sampling mechanism violates the i.i.d. assumption. Our methodology involves a formulation of the bias between the standard Cross-Validation (CV) estimator and the mean generalization error, denoted by $w_{cv}$, and practical data-based procedures to estimate this term. This concept was introduced in the literature only in the context of a linear model with squared error loss as the criterion for prediction performance. Our proposed bias-corrected CV estimator, $\text{CV}_c=\text{CV}+w_{cv}$, can be applied to any learning model, including deep neural networks, and to a wide class of criteria for prediction performance in regression and classification tasks. We demonstrate the applicability of the proposed methodology in various scenarios where the data contains complex correlation structures (such as clustered and spatial relationships) with synthetic data and real-world datasets, providing evidence that the estimator $\text{CV}_c$ is better than the standard CV estimator. This paper is an expanded version of our published conference paper$^{\text{1}}$.

\footnotetext[1]{Oren Yuval, Saharon Rosset. "Cross Validation for Correlated Data in Classification Models." The 28th International Conference on Artificial Intelligence and Statistics, 2025.} 
\end{abstract}

\noindent%
\vfill

\newpage
\spacingset{1.5}

\section{Introduction}\label{C4_Intro}
In many applications of statistical learning, the observations collected to train a supervised model are not sampled using an i.i.d. mechanism. Instead, the observations adhere to some correlation structure due to the nature of the data collection. The correlation between observations can arise from spatiotemporal relationships within the samples, as is common in neuroscience and ecology \cite{roberts2017cross}. It can also capture clustering of observations into groups, as is common, for example, in electronic medical records (same hospital, same city) and in computer vision (images of the same person). Within predictive modeling contexts, the presence of such correlation structures can influence model construction methodologies, resulting in a diverse array of approaches that span both traditional and modern techniques.

A simple example for this kind of violation of the i.i.d. sampling is when multiply samples are collected from each object, resulting in a correlation between them. A standard modeling method to address this kind of correlation structure is linear mixed models (LMM, \cite{verbeke1997linear}) with random intercept for grouped data:
\begin{align}\label{RandInter}
    y_{j,l}=x_{j,l}^t\beta+\delta_j+\epsilon_{j,l}  \hspace{1mm},\hspace{2mm} j\in \{1,...,J\}\hspace{1mm},\hspace{2mm} l \in \{1,...,n_j\},
\end{align}
where the index $(j,l)$ refers to the $l$th observation of object $j$, $x\in \mathbb{R}^p$ are the observed fixed effect covariates, $\beta \in \mathbb{R}^p$ is the fixed coefficients vector, $\delta_j$ is the random intercept of object $j$, the $\epsilon$'s are the iid errors, and $y$ is the observed outcome. In this setting, observations $y_{j,l_1}$ and $y_{j,l_2}$ are correlated because they share the same realization of $\delta_j$. This kind of statistical model is named a \textit{mixed model}, because it is an integration of the \textit{fixed effect}, which is characterized by the fixed vector $\beta$, and the \textit{random effect}, specified by the random vector $\delta$.   

Some other modeling methods are discussed in the traditional literature for dealing with correlated data, such as Gaussian process regression (GPR, \cite{williams1995gaussian}),  generalized least squares (GLS, \cite{hansen2007generalized}, and generalized linear mixed models (GLMM, \cite{stroup2012generalized} ). These methods have been shown to be practical in various fields (\cite{jiang2007linear}), such as medicine (\cite{brown2014applied}), geo-statistics \cite{goovaerts1999geostatistics}, genetic (\cite{maddison1990method}), and ecology (\cite{roberts2017cross}). In recent years, some studies \citep{simchoni2021using,tran2020bayesian,xiong2019mixed} suggested combining the idea of random effects with Deep Neural Networks (DNNs), and showed promising results. These approaches involve advanced methods to solve the optimization problem that takes into account the correlation structure, while maintaining the fitting procedure feasible in terms of computational resources. With any of the methods described above, one can be interested in using the fitted model to \textit{predict} a future outcome of the dependent variable. 

Given a statistical-based prediction task, it is of interest to evaluate the prediction error of the model in use. Usually, the model selection and tuning procedures are based on prediction error estimation, thus, it is important to obtain a reliable estimate. A widely used prediction error measure is the $K$-fold cross-validation (CV), (\citep{stone1974cross,hastie2009elements}), which is also the focus of this study. However, as discussed in the literature \citep{roberts2017cross,anderson2019comparing}, CV estimation might be biased and not suitable when the assumption of i.i.d. sampling is violated. Intuitively, if the correlation structure among training data differs from that between training data and prediction points, regular CV may inaccurately assess the generalization error, which is the expected loss of a learning process, in predicting a new realization of a prediction point. Traditionally, efforts to address this challenge of CV when dealing with changing correlation structures have focused on designing sampling approaches that would allow cross validation to give valid scores for model evaluation and selection, exactly or approximately, such as the leave cluster out (LCO) approach \cite{rice1991estimating}, or leave observation from each cluster out (LOFCO,\cite{wu2002local}). However, these methods are only applicable to particular correlation structures and necessitate a balanced data design, where the number of folds is dictated by the data structure.

The paper \cite{rabinowicz2022cross} introduces a new perspective on the CV estimation in the presence of correlated data, mainly in the context of linear models with squared error loss. One major contribution of that paper was to precisely identify the conditions for the unbiasedness of the CV estimator, concluding that it can also be unbiased if the correlated structure is preserved in the prediction points. In addition, the paper presents a bias-corrected CV measure, denoted CV$_c$, that was shown to be suitable in various scenarios where CV is biased due to the correlations in the data. In this study, we extend the results and the methodology of \cite{rabinowicz2022cross} to correct the CV estimator from linear models with square loss, to general models and general prediction criteria. In particular, we provide a framework for evaluating and selecting models for classification tasks in the presence of correlation structures. We derive a surprisingly general expression for the bias of CV for a class of prediction quality scores, independent of the model in use, and demonstrate that it can serve to explicitly describe this bias when using the most common measures of classification prediction error: 0-1 loss, cross entropy (Bernoulli log-likelihood) and hinge loss. The expressions we derive are simple and completely general, in that they do not depend on the true model or the modeling approach used for building the models. However, quantifying these general expressions in any specific setting requires knowledge of the model form and/or the model parameters. One approach we present for estimating this bias in practice is based on parametric bootstrap and is widely applicable; however, it can be computationally prohibitive in realistic settings. Therefore, we also propose several efficient approximations that apply to specific model settings, in particular when the true model is a GLMM. We also demonstrate that these methods can be used to approximately estimate the bias of the CV estimator in modern deep learning models and to better evaluate the prediction error. This paper is an expanded version of the published conference paper \cite{yuval2025cross}. The corresponding code is available at {\tt https://github.com/OrenYuv/CVc-for-General-Models}.

To make our overall findings both clear and concise, we concentrate on two main scenarios involving correlated data. 
\begin{enumerate}
    \item The multiple cluster structure with random intercepts model, which is discussed in \cite{gelman2007data}. For instance, repeatedly sampling the same entities on different days results in two observations being correlated if they come from the same entity or the same day. In this scenario, we might be interested in predicting future observations of an already observed entity (partial correlation), or in predicting observations from new, unsampled entities (zero correlation). 
    \item The scenario of spatial correlation structure (random fields), which occurs when data are collected from a geographical surface and the correlation between two observations is influenced by their spatial distance. For example, data are collected on real estate properties situated in particular geographic locations, and we are interested in predictions on properties in different locations that have a weaker correlation to the collected data.
\end{enumerate}

\subsection{Notations and assumptions}
We consider a statistical learning process that uses the training set $T=(X,Y)=\{x_i,y_i\}_{i=1}^n$ in order to be able to accurately predict the outcome $y_{te}$ by the fitted function $\hat{y}(x_{te})$ for a new covariate variable $x_{te}\in \mathbb{R}^p$, where $X\in \mathbb{R}^{n \times p}$ denotes the matrix of fixed effect covariates. Regarding the conditional distribution $P_{y|x}$, we can consider the following general setting, which covers a wide variety of models in the context of mixed-effect modeling in the literature.
\begin{align}\label{DeepGLMMForm}
    y_i|x_i,\delta_T \sim \mathcal{F}\left(g\left(f(x_i)+\delta_{T,i} \right) , \varphi \right),
\end{align}
where $\mathcal{F}$ is some distribution of the exponential family, with mean $g\left(f(x_i)+\delta_{T,i} \right)$ and a dispersion parameter $\varphi$. The vector $\delta_{T}\in \mathbb{R}^{n}$ is the \textit{random effect} vector that depends on the sampling mechanism and the realization of the latent variable that induces correlation across the $\delta_{T,i}$'s. The set of parameters of the distribution $P_{\delta}$, is denoted by $\gamma_r$. The \textit{fixed effect} function $f:\mathbb{R}^{p}\to \mathbb{R}$ is a nontrivial function for which DNNs and trees are suitable, with parameters $\gamma_c$. The link function $g:\mathbb{R}\to (0,1)$ is a monotonically increasing mapping between the mixed effect (fixed + random) and the expected value of the outcome. We also assume that the covariate vector $x_{te}$ is sampled from $P_x$, and the response $y_{te}$ drawn as follows:
\[y_{te}|x_{te} , \delta_{te} \sim \mathcal{F}\left(g\left( f(x_{te})+ \delta_{te}  \right) , \varphi \right),\]
where $\delta_{te}\in \mathbb{R}$ might depend on the entries of $\delta_T$ differently from how they depend on each other, due to the sampling mechanisms of $T$ and $(x_{te},y_{te})$. The outlined model covers the GLMM framework, taking $f(x)=x^t\beta$, where $\beta \in \mathbb{R}^{p}$ denotes an unknown vector of fixed coefficients. It also includes more complex mixed-effects models recently described in \cite{simchoni2023integrating}, \cite{xiong2019mixed}, and \cite{rabinowicz2022tree}. However, for simplicity, we assume that the prediction function $\hat{y}(x_{te};T)$ is deterministic in both $T$ and $x_{te}$, which means that there is no additional randomness in the learning process (such as that induced by using stochastic gradient descent in the training procedure). 

Considering the multiple cluster structure that involves $q_1$ entities and $q_2$ days, each entity $j_1\in \{1, \ldots, q_1\}$ generates a random intercept $u_{j_1} \sim N(0,\sigma_u^2)$, while each day $j_2\in \{1, \ldots, q_2\}$ generates a random intercept $s_{j_2}\sim N(0,\sigma_s^2)$. If the outcome $y_{i}$ is observed from the entity $j_1$ on day $j_2$, we can denote $i = j_1 ,j_2 ,l$, with $ l \in \{1,...,n_{j_1 ,j_2}\}$, where $n_{j_1 ,j_2}$ represents the total number of observations from the cluster $(j_1,j_2)$. The conditional distribution of this observation is determined by the additive random effect $\delta_{T,i}=u_{j_1} + s_{j_2}$. Usually, in the learning procedure, an estimator $\hat{\gamma}_c$ of $\gamma_c$ (the parameters of the fixed effect function) and predictors $(\hat{u},\hat{s}$ of the realizations $(u,s)$ are fitted based on the training data in hand. If the test pair $(x_{te},y_{te})$ is sampled from entity $j_1$ on a new day, the random effect $u_{j_1}$, is replicated, and we can write $\delta_{te}=u_{j_1}+ s_{te}$ where $s_{te}$ is a new realization, independent of the $s_{j_2}$'s. Then, it is customary to predict the outcome of $y_{te}$ by the estimated mean $\widehat{\mathbb{E}[y|x_{te}]}$ with $\gamma_c=\hat{\gamma}_{c}$, $u=\hat{u}$ and $s=0$, as there is no knowledge about the new random effect $s_{te}$, as follows: 
\[\hat{y}(x_{te};T)=g\left(f_{\hat{\gamma}_c}(x_{te})+\hat{u}\right).\] 
If $(x_{te},y_{te})$ is sampled from an unseen entity on a new day, we have $\delta_{te}=u_{te}+ s_{te}$ where $u_{te}$ is also a new realization independent of the $u_{j_1}$'s. In this scenario, predict the estimated mean with also $u=0$, as there is no knowledge about the new random effect $\delta_{te}$, as follows:
\[\hat{y}(x_{te};T)=g\left(f_{\hat{\gamma}_c}(x_{te})\right).\]

In the case of data with spatial Gaussian random field (GRF, \cite{kyriakidis1999geostatistical}) structure, the covariance between two observations is defined by a kernel function $\mathcal{K}(z_{i_1},z_{i_2}):\mathbb{R}^2 \times \mathbb{R}^2 \to \mathbb{R}^+$, which operates on pairs of coordinates. Let $K_{tot}\in \mathbb{R}^{(n+1) \times (n+1)}$ represent the matrix of the values of the kernel function between two coordinates, $K_{tot,i_1,i_2}=\mathcal{K}(z_{i_1},z_{i_2})$, $i_1,i_2 \in \{1, \cdots, n+1\}$, and let $\delta_{tot}^t=(\delta_{T}^t,\delta_{te})$ denote the random effect vector of the training and test data. In GRF we assume that $\delta_{tot}\sim MN(0_{n+1},K_{tot})$, where $MN$ stands for multivariate Gaussian distribution. In this scenario, we consider a specific situation in which training data is collected from $q$ geographical regions, randomly selected, and the test observation $(x_{te},y_{te})$ is sampled from another randomly chosen region. This typically results in stronger correlations between two observations within the same region compared to those between observations from different regions. In this scenario, incorporating a predictor to account for the specific realization of the random effect during the training phase is not a standard or intuitive approach. Thus, it is customary to predict the outcome of $y_{te}$ by $\hat{y}(x_{te};T)=g\left(f_{\hat{\gamma}_c}(x_{te})\right)$.

To assess the predictive capacity of a specific learning model, compare models, and tune hyperparameters, it is common to use the $K$-fold cross-validation procedure. This procedure uses a loss function $L(\cdot,\cdot): \mathbb{R} \times \mathbb{R} \to \mathbb{R}$, which quantifies the quality of the prediction. The procedure can be described as follows: The set $T$ is randomly ordered and divided into $K$ folds of size $\tilde{n}=n/K \in  \mathbb{N} $, $T_1=\{x_i,y_i\}_{i=1}^{\tilde{n}} , \cdots , T_K=\{x_i,y_i\}_{i=(K-1)\tilde{n}+1}^{n}$.  For each fold $k\in \{1, \cdots , K\}$, the model is trained on the set $T_{-k}=(X_{-k},Y_{-k}) = T\setminus T_k \in \mathbb{R}^{(n-\tilde{n}) \times (p+1)}$ which is the whole data without the observations of the $k$th fold. Then, the predicted outcome is calculated on the observations in the $k$th fold, $i \in \{(k-1)\tilde{n}+1,\cdots, k\tilde{n}\}$, and stored in the CV vector:
\[Y^{cv}_i=\hat{y}^{cv}_i :=\hat{y}(x_{i};T_{-k}).\]
The CV estimator is calculated by averaging the loss on the CV predictions:
\[\text{CV}=\frac{1}{n}\sum_{i=1}^{n}L\left(y_{i}, \hat{y}^{cv}_i \right).\]

In what follows, we use the notation $T_{-i}$ instead of $T_{-k}$ to denote the subset of $T$, of size $n-\tilde{n}$, without observations from the fold of the $i$th observation. We note that due to the randomness in the splitting process, any fold $T_{-i}$ shares the same distribution. Moreover, since any covariate vector from $(X,x_{te})$ is an i.i.d. sample from $P_x$, for any $i\in \{1,\cdots, n\}$, the pair $(T_{-i},x_i)$ shares the same distribution as $(T_{-1},x_{te})$.

The goal of the CV process is to estimate the generalization error (GenErr) of a modeling procedure (like a DNN with specific structure), which is the mean out-of-sample loss of this modeling approach applied to an arbitrary training set of size $n-\tilde{n}$, denoted by $T_{-1}$:
\begin{align}
    \text{GenErr}= \mathbb{E}_{\delta_{tot},T,x_{te},y_{te}} \hspace{1mm} L\left(y_{te}, \hat{y}(x_{te};T_{-1}) \right).
\end{align}
Using a similar notation and setting, the analysis in \cite{rabinowicz2022cross} found that the CV estimator may be a biased estimator of GenErr when the correlations between $y_{te}$ and $T_{-1}$ differ compared to the correlation between $y_1$ and $T_{-1}$. However, the discussion in \cite{rabinowicz2022cross} was limited to regression settings, where the loss function $L$ is the square difference between $y_{te}$ and $\hat{y}(x_{te};T_{-1})$. We show next that the criterion for CV unbiasedness is the same as presented in \cite{rabinowicz2022cross}, without any assumption on the loss function $L$. In the sequel, we define a general form of $L$, which covers the squared loss and many others, inducing prediction criteria for binary classification tasks, under which we can explicitly express the bias of the CV estimator with respect to the generalization error, denoted by $w_{cv}$. The correlation-corrected CV estimator is subsequently defined as: \[\text{CV}_c=\text{CV}+w_{cv}.\]
Later, we suggest an empirical evaluation of the adjusted CV version by employing a parametric bootstrap technique, suitable for complex non-linear models, for estimating the bias term $w_{cv}$.

\section{The bias of CV with general loss function}
\subsection{Criterion for CV Unbiasedness}
The main idea in correcting the CV estimator, as presented in \cite{rabinowicz2022cross}, is to identify the bias of the estimator with respect to the generalization error, denoted by $w_{cv}$, and to correct the CV by adding some estimate of $w_{cv}$. Under the assumption of squared loss, the following formulation of $w_{cv}$ was presented:
\[w_{cv}=\frac{2}{n}\text{tr} \left(\mathbb{E}_X\left[ \mathbb{C}ov\left( Y^{cv},Y |X\right) \right] \right)- 2\mathbb{E}_{X_{-1},x_{te}}\left[ \mathbb{C}ov\left( \hat{y}(x_{te};T_{-1}) ,y_{te} |X_{-1},x_{te}\right) \right].\]
Using the above expressions, it was shown that $w_{cv}=0$ when $P_{x_{te},y_{te},T_{-1}} = P_{x_i,y_i,T_{ - i}}$ $\forall i \in  \{1,\cdots , n\}$. However, we can show that this is a sufficient condition, regardless of the definition of the loss function. Because of the random ordering of $T$, the observation $\{x_i,y_i\}$ are all exchangeable, and follow the same joint distribution, and together with the property  $P_{x_{te},y_{te},T_{-1}} = P_{x_1,y_1,T_{ - 1}}$ we get that :
\begin{align}\label{UnbiasCrit}
    \mathbb{E}[\text{CV}] &= \mathbb{E}\frac{1}{n}\sum_{i=1}^{n}L\left(y_{i}, \hat{y}^{cv}_i \right) =\mathbb{E}\frac{1}{n} \sum_{i=1}^{n} \hspace{1mm} L\left(y_i, \hat{y}(x_i;T_{ - i}) \right) =  \mathbb{E}  \hspace{1mm} L\left(y_1, \hat{y}(x_1;T_{ - 1}) \right) \\
    &= \mathbb{E} \hspace{1mm} L\left(y_{te}, \hat{y}(x_{te};T_{-1}) \right) = \text{Generalization error}.
\end{align}

This intuitive result is important to avoid the misconception that the CV estimator is biased in any case of correlation within the training data. In particular, if the same correlation (induced by the same random effect) is preserved between the test observations and the training set, the standard CV estimator is unbiased and therefore suitable. We also note that in our general perspective, there is no binding relationship between the statistical model and the loss function $L$. For example, the model can be non-linear, and the loss function can still be the squared error.

\subsection{Formulation of CV bias for general loss function}
We now present a specific structure, but yet very general, of the loss function, under which we can obtain an interpretable formulation of $w_{cv}$, while we cover many practical scenarios. We consider that the loss function, $L(y, \hat{y})$, can be written as follows:
\begin{align}\label{LossDef}
    L\left(y, \hat{y} \right) = L_1\left(\hat{y} \right) - L_{2}\left(\hat{y} \right) \cdot y +L_3(y),
\end{align}
where each of the functions $L_1$, $L_{2}$, and $L_{3}$ is an arbitrary function on the real space. We can see that in the special case of squared error, we have $L_1(y)=L_3(y)=y^2$, and $L_{2}(\hat{y})=2\hat{y}$. 

Using the structure of $L$, together with the definition of $w_{cv}$ as the difference between the generalization error and the expected value of CV, we can derive the following expressions:
\begin{align} \nonumber
    w_{cv}&= \text{Generalization error} - \mathbb{E}[\text{CV}]  = \mathbb{E} \hspace{1mm} L\left(y_{te}, \hat{y}(x_{te};T_{-1}) \right) - \frac{1}{n}\sum_{i=1}^{n} \mathbb{E} \hspace{1mm} L\left(y_i, \hat{y}^{cv}_i \right)\\ \nonumber
    &= \mathbb{E}\left[ L_1\left(\hat{y}(x_{te};T_{-1}) \right) - L_{2}\left(\hat{y}(x_{te};T_{-1})\right) \cdot y_{te}  +L_3(y_{te}) \right]   \\ \nonumber & - \frac{1}{n}\sum_{i=1}^n \mathbb{E} \hspace{1mm} \left[ L_1\left(  \hat{y}^{cv}_i  \right) \hspace{2mm} - \hspace{2mm} L_{2}\left( \hat{y}^{cv}_i\right) \cdot y_{i}  \hspace{3mm}+ \hspace{3mm} L_3(y_{i})\right],
\end{align}
and since the $y_i$ and $y_{te}$ have the same marginal distribution, the expectations $ \mathbb{E}\left[ L_3(y_{te}) \right] $ and $ \mathbb{E}\left[ \frac{1}{n}\sum_{i=1}^{n} L_3(y_{i}) \right] $ cancel in the above expression. Moreover, since $(x_i,T_{-i})$ have the same distribution as $(x_{te},T_{-1})$, we conclude that $\hat{y}^{cv}_i$ and $\hat{y}(x_{te};T_{-1})$ have the same marginal distribution, and the expectations $ \mathbb{E}\left[ L_1(\hat{y}(x_{te};T_{-1})) \right] $ and $ \mathbb{E}\left[ \frac{1}{n}\sum_{i=1}^{n} L_1(\hat{y}^{cv}_i ) \right] $ also cancel. Therefore we can write:
\begin{align} \nonumber
    w_{cv}&= \frac{1}{{n}}\sum_{i=1}^{n} \mathbb{E} \left[ L_{2}\left( \hat{y}^{cv}_i\right) \right] \cdot \mathbb{E} \left[y_{i}\right] + \mathbb{C}ov\left( L_{2}\left( \hat{y}^{cv}_i\right) , y_{i}\right) \\ \nonumber & -  \mathbb{E}\left[ L_{2}\left(\hat{y}(x_{te};T_{-1})\right) \right] \cdot \mathbb{E}\left[y_{te} \right] -  \mathbb{C}ov\left( L_{2}\left( \hat{y}(x_{te};T_{-1})\right) , y_{te}\right)\\ \label{WcvForm1}
    &= \frac{1}{{n}}\sum_{i=1}^{n} \mathbb{C}ov\left( L_{2}\left( \hat{y}^{cv}_i\right) , y_{i}\right) - \mathbb{C}ov\left( L_{2}\left( \hat{y}(x_{te};T_{-1})\right) , y_{te}\right) 
\end{align}

We note that the pair $(X_{-i},x_i)$ and $(X_{-1},x_{te})$ are identically distributed, and for any realization $(\widetilde{X},\tilde{x})$, the following equations holds true:
\begin{align*}
    \mathbb{E}[y_{te}|x_{te}=\tilde{x}] &=\mathbb{E}[y_{i}|x_{i}=\tilde{x}], \\
    \mathbb{E}\left[L_{2}\left( \hat{y}(x_{te};T_{-1})\right)|(X_{-1},x_{te})=(\widetilde{X},\tilde{x})\right] &=\mathbb{E}\left[L_{2}\left( \hat{y}^{cv}_i\right)|(X_{-i},x_i)=(\widetilde{X},\tilde{x})\right].
\end{align*}
Therefore, when we apply the law of total covariance, the $\mathbb{C}ov\left(\mathbb{E}\left[ L_{2}\left( \hat{y}^{cv}_i\right)|X  \right] , \mathbb{E} \left[ y_{i} | x_i \right] \right) $ term is cancel out with the term $\mathbb{C}ov\left( \mathbb{E}\left[L_{2}\left( \hat{y}(x_{te};T_{-1})\right) | X_{-1},x_{te} \right], \mathbb{E}\left[y_{te} | x_{te} \right]\right) $, and we derive the final result:
\begin{align} \label{WcvForm2}
w_{cv} &= \frac{1}{{n}}\sum_{i=1}^{n} \mathbb{E}_X\left[ \mathbb{C}ov_{\delta_T,Y}\left(L_{2}\left( \hat{y}^{cv}_i\right) , y_{i} | X  \right) \right] \\ \nonumber &  - \mathbb{E}_{X_{-1},x_{te}}\left[ \mathbb{C}ov_{\delta_{T_{-1}},Y_{-1},\delta_{te},y_{te}}\left(L_{2}\left( \hat{y}(x_{te};T_{-1})\right) ,y_{te} | X_{-1},x_{te} \right) \right].
\end{align}
From the above expression, we conclude that, in general, the bias of the CV estimator is equivalent to the extra covariance between $L_{2}(\hat{y}^{cv}_i)$ and $y_i$ over the one of $L_{2}(\hat{y}(x_{te};T_{-1}))$ and $y_{te}$. This is our key result, and it applies to a wide class of loss functions, such as squared loss for regression, and in that sense it generalizes the previous results by \cite{rabinowicz2022cross}.

A special case of the presented setting is when $\delta_{te}$ is completely independent of $\delta_{T_{-1}}$. In this case, conditionally on $(X_{-1},x_{te})$, the outcome $y_{te}$ is independent of $Y_{-1}$ and also of $\hat{y}(x_{te};T_{-1})$, and we see that the last term in the formulation of $w_{cv}$ in Equation~(\ref{WcvForm2} )is equal to zero. This situation arises, for example, in the multiple cluster structure scenario where $y_{te}$ is sampled from an unseen entity on a new day, and therefore $\delta_{te}=u_{te}+s_{te}$ is completely independent of $\delta_T$.


The form of the loss function $L\left(y, \hat{y} \right)$ might seem arbitrarily and not very practical at first sight. However, it covers the case where $L$ corresponds to the negative log likelihood of natural parameter, $\theta$, of the distribution $\mathcal{F}$, given the observed value $y_{te}$.  In the distributional model described in (\ref{DeepGLMMForm}), let us denote the conditional mean by $\mu=g\left( f(x_{te})+ \delta_{te}  \right)$, then, the natural parameter of the distribution $\mathcal{F}$ can be written as $\theta=L_{2}(\mu)$ for some monotonically increasing function $L_{21}$, and the density $p(y|\theta)$ satisfies the following.
\begin{align}\label{NLLProp}
    -log\left(p(y) \right) \propto  G(\theta)-  \theta y = G(L_{2}(\mu))-  L_{2}(\mu) y, 
\end{align}
where $G$ is a convex function satisfies $G'(\theta)=L_{2}^{-1}(\theta)=\mu$. Now, we recall that $\hat{y}$ represents an estimated value of $\mu$, and therefore the criterion $L(y,\hat{y})=   G\left( L_{2}(\hat{y})\right)   - \hspace{1mm}L_{2}(\hat{y}) \cdot y$ is proportional the negative log likelihood of the natural parameter that determined by $\hat{y}$. We conclude that the suggested form of $L$ is useful in many scenarios where the criterion for model performance is based on the distributional assumptions on the outcome $y$. In the next subsection, we show the applicability of this form of the loss function to binary classification, where many different metrics can be used for model evaluation and selection.

\subsection{Application to binary classification}
In a classification task, the loss function $L(y,\hat{y}): \{0,1\}\times \mathbb{R}\to \mathbb{R}$ assesses the discrepancy between the actual response variable $y\in \{0,1\}$ and the predicted value $\hat{y}$. We point to the fact that any such function $L(y,\hat{y})$ can be decomposed as follows:
\begin{align}\nonumber
    L(y,\hat{y}) &=y\cdot L(1,\hat{y}) + (1-y) L(0,\hat{y})  \\ \label{LStructure}
    &= \underset{L_{1}(\hat{y})}{\underbrace{ L(0,\hat{y})}}      - \hspace{1mm}\underset{L_{2}(\hat{y})}{\underbrace{\left[L(0,\hat{y})-L(1,\hat{y}) \right]}} \cdot y .
\end{align}
Therefore, according to formula~(\ref{WcvForm1}),  the bias of the CV estimator can be identified as the following difference between the covariances
\[\mathbb{C}ov\left( L\left(0,\hat{y}^{cv}_1\right) - L\left(1,\hat{y}^{cv}_1\right), y_{1}\right)  - \mathbb{C}ov\left( L\left(0, \hat{y}(x_{te};T_{-1})\right)- L\left(1, \hat{y}(x_{te};T_{-1})\right) , y_{te}\right).\]

This can be illustrated in standard classification settings:
\begin{enumerate}
    \item The output $\hat{y} \in (0,1)$ aims to determine the probability that $y=1$, and the performance is measured by the cross entropy loss. In this case, we can write:
    \begin{align*}
        L(y,\hat{y})&= - y\cdot log(\hat{y}) -(1-y)\cdot log(1-\hat{y}) \\&=
        \underset{L_{1}(\hat{y})}{\underbrace{- log(1-\hat{y})}}    - \hspace{1mm}\underset{L_{2}(\hat{y})}{\underbrace{log\left(\frac{\hat{y}}{1-\hat{y}}\right)}} \cdot y.
    \end{align*}
    \item The output $\hat{y} \in \{0,1\}$ is the best guess for the actual value of $y$, and the performance is measured by the zero-one loss. We can express this as follows:
\[L(y,\hat{y})=  y\cdot(1-\hat{y})+ (1-y)\cdot\hat{y}   = \underset{L_{1}(\hat{y} )}{\underbrace{\hat{y} }} \hspace{1mm}- \hspace{1mm} \underset{L_{2}(\hat{y})}{\underbrace{(2\hat{y}-1) }} \cdot y.\]
    \item The output $\hat{y} \in \{-\infty,\infty\}$ is some score in favor of $y=1$, and the performance is measured by the Hinge loss. In this case, the penalty is zero if $y=1$ and $\hat{y}\geq 1$, or $y=0$ and $\hat{y}\leq -1$, and increases linearly in $\hat{y}$ otherwise. This can expressed as follows:
\begin{align*}
    L(y,\hat{y}) &=  y\cdot \text{ReLU}(1-\hat{y})+ (1-y)\cdot \text{ReLU}(1+\hat{y})\\
    &= \underset{L_{1}(\hat{y})}{\underbrace{\text{ReLU}(1+\hat{y}) }}  -\underset{L_{2}(\hat{y})}{\underbrace{  \left[\text{ReLU}(1+\hat{y}) -\text{ReLU}(1-\hat{y})\right]     }}  \cdot y ,
\end{align*}
where ReLU stands for the rectified linear unit, $\text{ReLU}(x)=max\{0,x\}$.
\end{enumerate}
Taking into account the analysis and examples provided, we can infer that our proposed formulation for the bias term $w_{cv}$ is entirely appropriate for utilization in classification tasks. Furthermore, we reiterate the observation that there is no inherent dependency between the design of the model's architecture and the metric used for its evaluation.

Another important performance measurement of a binary classifier is the 'Area under the ROC curve' (AUC, for short). This metric exhibits greater complexity compared to any conventional loss function that is averaged across the available data set. However, a comparable methodology can be utilized to suitably amend this measure, thereby allowing it to incorporate the correlations present within the data set. In general, the $\widehat{AUC}$ statistic is calculated on the input $(\widehat{\text{FPR}},\widehat{\text{TPR}})\in \mathbb{R}^{n_t \times 2}$ where $\widehat{\text{FPR}}$ ($\widehat{\text{TPR}}$) is an estimation of the False (True) Positive Rate  with different threshold values $(\zeta_1,\cdots \zeta_{n_t})^t$. In the case where the $\widehat{\text{FPR}}$ statistic tends to underestimate the real FPR and the $\widehat{\text{TPR}}$ statistic tends to overestimate the real TPR, the $\widehat{AUC}$ statistic will be too optimistic compared to the real performance. Although we cannot express the AUC measure as a loss function in the suggested pattern, we can find a bias correction term to the classical $(\widehat{\text{FPR}},\widehat{\text{TPR}})$ estimates. Then we calculate the approximate corrected statistic $\widehat{AUC}_c$ based on the corrected estimates $(\widehat{\text{FPR}}_c,\widehat{\text{TPR}}_c)$.

For a given threshold value $\zeta$, the output $\hat{y}$ is labeled by $1$ if the predicted value for the probability that $y=1$ is greater than $\zeta$. Denoting by $\hat{y}^{cv}_i(\zeta)$ the predicted $0/1$ label of observation $i$ in the CV procedure, with threshold value $\zeta$, and by $n_0$ the number of observations satisfying $y=0$, we can write:
\[\widehat{\text{FPR}}(\zeta) = \frac{1}{n_0}\sum_i (1-y_i)\hat{y}^{cv}_i(\zeta) = \frac{n}{n_0} \cdot \frac{1}{n}\sum_i (1-y_i)\hat{y}^{cv}_i(\zeta). \]
On the other hand, the real FPR of the model can be written as:
\begin{align*}
    \text{FPR}(\zeta)&=P(\hat{y}(x_{te};T_{-1},\zeta)=1|y_{te}=0)= \frac{P(\hat{y}(x_{te};T_{-1},\zeta)=1 , y_{te}=0)}{P(y_{te}=0)}\\
    &:= \frac{1}{p_0} \cdot \mathbb{E}\left[(1-y_{te})\hat{y}(x_{te};T_{-1},\zeta)\right],
\end{align*}
where $p_0$ is the marginal probability of $y=0$.

Based on the previous analysis, we can see that the bias between the right term of $\widehat{\text{FPR}}(\zeta)$ and the right term of FPR, $p_0\text{FPR}(\zeta) - \mathbb{E} \left[\frac{n_0}{n}\widehat{\text{FPR}}(\zeta)\right] $ is exactly:
\[w_{cv}^{PR}(\zeta):=\frac{1}{{n}}\sum_{i=1}^{n} \mathbb{E}_X\left[ \mathbb{C}ov\left(\hat{y}^{cv}_i(\zeta) \hspace{1mm} ,\hspace{1mm} y_{i} |X\right) \right] - \mathbb{E}_{X_{-1},x_{te}}\left[ \mathbb{C}ov\left( \hat{y}(x_{te};T_{-1},\zeta) ,y_{te} | X_{-1},x_{te} \right) \right].\]
The above result allows us to introduce the following adjusted estimator for the true FPR:\[\widehat{\text{FPR}}_c(\zeta) = \widehat{\text{FPR}}(\zeta)+\frac{n}{n_0}w_{cv}^{PR}(\zeta). \]
In the same manner, we can define a corrected estimator for the real TPR as follows:
\[\widehat{\text{TPR}}_c(\zeta) = \widehat{\text{TPR}}(\zeta)-\frac{n}{n_1}w_{cv}^{PR}(\zeta), \]
where $n_1$ is the number of observations satisfying $y=1$. 

Assuming that the triplet $(\widehat{\text{FPR}},\widehat{\text{TPR}},\zeta)\in \mathbb{R}^{n_t \times 3}$ is given by standard computational procedure, we can find a corrected measures $(\widehat{\text{FPR}}_c,\widehat{\text{TPR}}_c)$ such that:
\begin{align*}
    \widehat{\text{FPR}}_{c,l}&=\widehat{\text{FPR}}_l+\frac{n}{n_0}w_{cv}^{PR}(\zeta_l) \hspace{2mm};\hspace{2mm} l=1\cdots n_t,\\
    \widehat{\text{TPR}}_{c,l}&=\widehat{\text{TPR}}_l-\frac{n}{n_1}w_{cv}^{PR}(\zeta_l) \hspace{2mm};\hspace{2mm} l=1\cdots n_t.
\end{align*}
Given $(\widehat{\text{FPR}}_c,\widehat{\text{TPR}}_c)\in \mathbb{R}^{n_t \times 3}$, we can calculate a corrected measure $\widehat{AUC}_c$ that takes into account the correlations in the training set, using a standard computational procedure.

In practical applications, it is necessary to estimate the terms $w_{cv}$ and $w_{cv}^{PR}(\zeta)$ based on the available data in order to assess a refined version of the CV estimator and the AUC metric. The following section addresses various approaches for evaluating these estimates.

\section{Methods for estimating the bias of the standard CV estimator}\label{MethodsForEst}
We suggest a parametric bootstrap procedure in which we utilize the covariate matrix at hand, $X$, and estimates $\hat{\gamma}_c$ and $\hat{\gamma}_r$ of the true parameters $\gamma_c$ and $\gamma_r$, which can be obtained by the training sample $T$. This approach does not require any prior knowledge about the distribution $P_x$ nor any distributional assumptions about it. However, it is necessary to simplify $w_{cv}$ from equation (\ref{WcvForm2}), according to the specific scenario. Our objective is to derive a formulation that excludes the expectation over $x_{te}$, which is not observed in the available training dataset.

In the scenario of the structure of multiple levels clusters, where the test observation is sampled from the same set of entities on a new day, let us denote by $u_{tr}\in \mathbb{R}^{q_1}$, the vector of the realizations of the random intercepts of the entities in the training data. Since the random intercept of the new day is independent of those of the training days, the following holds:
\[  \mathbb{C}ov\left(L_{2}\left( \hat{y}(x_{te};T_{-1})\right) , y_{te} | X_{-1},x_{te}, u_{tr}\right) = 0.\] 
Moreover, when conditioning on $u=u_{tr}, (X_{-1},x_{te})=(\widetilde{X},\tilde{x})$ and $(X_{-i},x_i)=(\widetilde{X},\tilde{x})$, we get equalities in the expectations:
\[ \mathbb{E}[y_i]=\mathbb{E}[y_{te}] \text{  and  } \mathbb{E}[L_{2}\left( \hat{y}(x_{te};T_{-1})\right)]=\mathbb{E}[L_{2}\left( \hat{y}^{cv}_i\right)].\] As a result, by applying the law of total covariance to Equation (\ref{WcvForm2}), we obtain the appropriate formula for $w_{cv}$ in this context:
\[w_{cv}=\frac{1}{{n}}\sum_{i=1}^{n} \mathbb{E}_{X,u_{tr}}\left[ \mathbb{C}ov\left(L_{2}\left( \hat{y}^{cv}_i\right) , y_{i}  | X,u_{tr} \right) \right].\]

The above result shows that, in a clustered correlation structure, the bias of the CV estimator is equal to the mean conditional covariance between the CV predictions, $L_2(y^{cv}_i)$'s and the labels $y_i$'s, when the conditioning yields independence between $y_{te}$ and $\hat{y}(x_{te};T_{-1})$. 

Assume that the covariate matrix $X=\{x_i\}_{i=1}^n$ is given in random order, and let us denote by $C_i$, the mean conditional covariance between $L_2(y^{cv}_i)$ and $y_i$: $C_i=\mathbb{E}_{u_{tr}}\mathbb{C}ov \left(L_{2}( y^{cv}_i),y_i  | u_{tr}\right)$. We note that each $C_i$ is an unbiased estimator of $w_{cv}$, and we suggest estimating $w_{cv}$ by averaging the estimates $C_i$ for $ i=1,\cdots , n$. Formally:
\[\hat{w}_{cv} = \frac{1}{n}\sum_i \widehat{C}_i.\]

Estimation of $C_i$ is done by the following parametric bootstrap procedure:

\begin{enumerate} 
    \item Evaluate estimates of the unknown parameters, denoted by $\hat{\gamma}_c$, $\hat{\gamma}_r=(\hat{\sigma}_u^2,\hat{\sigma}_s^2)$, and $\hat{\varphi}$.
    \item For $b_1=1,\cdots ,B_1$:
    \begin{enumerate} 
        \item Draw $u_{b_1}\sim MN(0_{q_1},\hat{\sigma}^2_{u}I_{q_1})$
        \item For $b_2=1,\cdots ,B_2$:
        \begin{enumerate} 
            \item  Draw $s_{b_2}\sim MN(0_{q_2},\hat{\sigma}^2_{s}I_{q_2})$
            \item  Draw $Y_{b_1,b_2}$ from the distribution $\mathcal{F}\left(g\left(f_{\hat{\gamma}_c}(X)+\delta(u_{b_1},s_{b_2}) \right) , \hat{\varphi} \right)$, where $f_{\hat{\gamma}_c}$ is the estimated fixed-effect function, and $\delta(u_{b_1},s_{b_2})\in \mathbb{R}^n$ with $\delta(u_{b_1},s_{b_2})_i=u_{b_1,j_1}+s_{b_2,j_2}$ if observation $i$ samples from entity $j_1$ on day $j_2$. \\ *In a classification framework, for example, $Y_{b_1,b_2}$ is drawn from the Bernoulli distribution with mean $g\left(f_{\hat{\gamma}_c}(X)+\delta(u_{b_1},s_{b_2}) \right)$.
            \item Apply the learning procedure to the drawn training set $T_{b_1,b_2,-i}=(X_{-i},Y_{b_1,b_2,-i})$, and calculate $l_{b_1,b_2,i}= L_{2}(\hat{y}(x_{i};T_{b_1,b_2,-i}))$.            
        \end{enumerate}

        \item  Calculate the inner-sample covariance:
        \[\widehat{C}_{i,b_1}  = \frac{1}{B_2}\sum_{b_2=1}^{B_2} (l_{b_1,b_2,i} - \overline{l_{b_1,i} })(y_{b_1,b_2,i}  - \overline{y_{b_1,i} }),\]
        \[ \overline{l_{b_1,i} }= \frac{1}{B_2}\sum_{b_2=1}^{B_2} l_{b_1,b_2,i}   \hspace{1mm};\hspace{1mm}  \overline{y_{b_1,i} }= \frac{1}{B_2}\sum_{b_2=1}^{B_2} y_{b_1,b_2,i}.  \]
    \end{enumerate}
    \item Calculate the mean: $\widehat{C_i}=\frac{1}{B_1}\sum_{b_1=1}^{B_1} \widehat{C}_{i,b_1}.$
\end{enumerate}

Following the basic properties of the bootstrap procedure, the result of the suggested algorithm converges to the quantity of interest, $C_i$, if $(\hat{\gamma}_c,\hat{\gamma}_r,\hat{\varphi})=(\gamma_c, \gamma_r,\varphi)$, and $B_1,B_2\to \infty$. However, the major downside is that it requires a great deal of computing power, especially at Step 2.b.iii, where we apply the learning procedure $B_1\cdot B_2$ times. In Section \ref{Fast-Approximated}, we show how this problem can be resolved by applying an approximate estimate of $l_{b_1,b_2,i}$, instead of the complete fitting process. 

In the scenario of spatial correlation, we are not able to achieve an exact simplified formula of $w_{cv}$ for a general model. Instead, we suggest to embrace a simplified formula that holds in the Linear Mixed models (LMM, \cite{galecki2013linear}) framework, as an approximate measure of $w_{cv}$. In LMM, the link function $g$ is identity, and the outcome vector $(Y_{-i},y_i)$ is linear in the random effect vector $(\delta_{-i},\delta_i)$. Moreover, as discussed in \cite{rabinowicz2022cross}, the prediction of $x_{i}$ is linear in $Y_{-1}$, and can be written as $\hat{y}(x_i;T_{-i})=h_i^tY_{-1}$ where the vector $h_i$ depends on the covariates $X_{-i}$ and $x_i$. Finally, when considering the squared loss criterion, $L_2(\hat{y})=2\hat{y}$, which can be taken outside the covariance operator, and we can write:
\[\mathbb{C}ov\left( \hat{y}^{cv}_i, y_{i} | X,K_{tot}  \right) =h_i^tK_{tot_{-i,i}} \hspace{1mm};\hspace{1mm} i\in\{1,\cdots n\}, \]
where $K_{tot_{-i,i}}$ is the vector of the covariance values between $\delta_i$ and each entry of $\delta_{-i}$. We note that the whole matrix $K_{tot}$ depends on the realization of the coordinates $Z=\{z_i\}_{i=1}^n$ and $z_{te}$, but not on $X$. Similarly, we can write:
\[\mathbb{C}ov\left(\hat{y}(x_{te};T_{-1}) ,y_{te} | X_{-1},x_{te} ,K_{tot} \right) = h_{te}^tK_{tot_{-1,te}},\]
where the vector $h_{te}$ depends on the covariates $X_{-1}$ and $x_{te}$, and $K_{tot_{-1,te}}$ is the vector of the values of covariance between $\delta_{te}$ and each entry of $\delta_{-1}$. Let us define $\tilde{k}_i =K_{tot_{-i,i}}-K_{tot_{-1,te}} $, the random vector of differences between the value of covariance between the sampling of $Z$ and $z_{te}$, and note that $h_i$ and $h_{te}$ are identically distributed, we can express $w_{cv}$ in the LMM scenario as follows:
\begin{align*}
    w_{cv}&=2\left(\mathbb{E}_{X,Z}\left[ h_i^tK_{tot_{-i,i}} \right] -\mathbb{E}_{X,Z,x_{te},z_{te}}\left[ h_{te}^tK_{tot_{-1,te}} \right]\right) \\ &=2\cdot\mathbb{E}\left[ h_i^t\tilde{k}_i\right] = \frac{2}{{n}}\sum_{i=1}^{n}\mathbb{E}\left[ \mathbb{C}ov\left( \hat{y}^{cv}_i , y_{i} | X ; \widetilde{K} \right) \right] = \frac{1}{{n}}\sum_{i=1}^{n}\mathbb{E}\left[ \mathbb{C}ov\left( L_2(\hat{y}^{cv}_i) , y_{i} | X ; \widetilde{K} \right) \right],
\end{align*}
where the notation $\mathbb{C}ov(\cdot,\cdot ; \widetilde{K})$ implies that the randomness of $Y$ is induced by random effect vector $\delta \sim MN(0,\widetilde{K})$ and each column of $\widetilde{K}$ satisfies $\widetilde{K}_i=\tilde{k}_i$.

The aforementioned formula enables an approximate estimation of the $w_{cv}$ term in the context of region-wise sampling, particularly within a general framework where the loss function $L$ adheres to the structure outlined in \ref{LossDef}. We do it by defining the covariance matrix $\widetilde{K}$ as follows: $\widetilde{K}_{i_1i_2}=0$ if $i_1$ and $i_2$ are not from the same region, and otherwise, $\widetilde{K}_{i_1i_2}= \widehat{\mathcal{K}}(z_{i_1},z_{i_2})-\overline{K}$, where $\widehat{\mathcal{K}}$ is the kernel function with the estimated parameters $\hat{\gamma}_r$, and $\overline{K}$ is the estimated mean covariance between the two observations from different regions. In particular, considering the Gaussian kernel function, the kernel parameters are $\sigma_{out}^2$ and $\sigma_{in}^2$, and we can write:
\[\widehat{\mathcal{K}}(z_{i_1},z_{i_2} ; \hat{\sigma}_{out}^2,\hat{\sigma}_{in}^2) = \hat{\sigma}_{out}^2 \cdot exp\{ - \hat{\sigma}_{in}^2 || z_{i_1}-z_{i_2}||_2^2 \},   \]
where $\hat{\sigma}_{out}^2$ and $\hat{\sigma}_{in}^2$ are estimators of $\sigma_{out}^2$ and $\sigma_{in}^2$, respectively.

The algorithm designed for clustered scenarios can be modified for this context as follows:
\begin{enumerate} 
    \item Evaluate estimates of the unknown parameters, denoted by $\hat{\gamma}_c$, $\hat{\gamma}_r=(\hat{\sigma}_{out}^2,\hat{\sigma}_{in}^2)$, and $\hat{\varphi}$.
    \item Evaluate the adjusted covariance matrix $\widetilde{K}=\widetilde{K}(Z;\hat{\sigma}_{out}^2,\hat{\sigma}_{in}^2)$.
    \item For $b=1,\cdots ,B$:
    \begin{enumerate} 
        \item Draw random effect vector $\delta_b$ from $MN(0,\widetilde{K})$.            
        \item  Draw $Y_{b}$ from the distribution $\mathcal{F}\left(g\left(f_{\hat{\gamma}_c}(X)+\delta_b \right) , \hat{\varphi} \right)$.
        \item Apply the learning procedure to the drawn training set $T_{b,-i}=(X_{-i},Y_{b,-i})$, and calculate $l_{b,i}= L_{2}(\hat{y}(x_{i};T_{b,-i}))$.            
        
    \end{enumerate}
    \item  Calculate the sample covariance $l_{i}$ and $y_{i}$:
        \[\widehat{C}_{i}  = \frac{1}{B}\sum_{b=1}^{B} (l_{b,i} - \overline{l_{i} })(y_{b,i}  - \overline{y_{i} }).\]
\end{enumerate}

The algorithms described for the estimation of $w_{cv}$ in the two scenarios discussed can be adjusted and extended to fit other forms of correlation structures. The main principle remains to derive a simplified expression of $w_{cv}$ based on formula~(\ref{WcvForm2}) and the specific statistical model. Then, the bootstrap procedure is a straightforward implementation. The problem of high computational resource demand is the same in all scenarios, but can be resolved by applying an approximated calculation of $\hat{y}(x_{i};T_{b,-i})$ (or $\hat{y}(x_{i};T_{b_1,b_2,-i})$) inside the bootstrap loop, as we describe in the next section.

\subsection{Fast-Approximated estimation for GLMM model}\label{Fast-Approximated}
In the GLMM framework, the assumed fixed model is $f(x)=x^t\beta$, and the statistical model for $y|x,\delta$, described in~\ref{DeepGLMMForm}, can be simplified to:
\begin{align}\label{GLMMForm}
    y\sim \mathcal{F}\left(g\left(x^t\beta+\delta\right) , \varphi \right).
\end{align}
Moreover, the density $p(y|x,\delta)$ is well defined by a coefficient vector $\beta$ and the dispersion parameter $\varphi$, which is set naturally at $1$ in some cases. For example, in a binary classification model, the distribution $P_{y|x,\delta}$ is Bernoulli, with success rate $g(x^t\beta+\delta)$ and $\varphi=1$, where $g$ is the Sigmoid function in the canonical case, and can be the CDF of the standard normal distribution in the non-canonical case.

For any given training fold $T_{-k}$, $k\in \{1,\cdots,K\}$, the learning procedure for fitting an estimator to $\beta$ involves the optimization of the integrated likelihood, $\mathcal{L}$ of the training sample $T_{-k}$, with respect to $(\beta,\varphi,\gamma_r)$, which can be written as follows:
\begin{align}\label{Likelihood1}
    \mathcal{L}= \int p(Y_{-i}|X_{-i}, \delta ;\beta,\varphi)p(\delta;\gamma_r)d\delta = \int \prod_{i=(k-1)\tilde{n}+1}^{k\tilde{n}} p(y_{i};x_{i},\beta,\varphi|\delta)p(\delta;\gamma_r)d\delta.
\end{align}
For example, in classification model for binary response, we have $\varphi=1$ by definition, and the fitting procedure of the estimators $(\hat{\beta},\hat{\gamma}_r)$,(as described in \cite{zhang2011fitting}), involves the optimization of the Bernoulli likelihood, $\mathcal{L}$ of a given training sample $T_{-k}$, $k\in \{1,\cdots,K\}$, with respect to $(\beta,\gamma_r)$, and can be written as follows:
\begin{align}\label{Likelihood2}
    \mathcal{L}= \int \prod_{i=(k-1)\tilde{n}+1}^{k\tilde{n}} g_i^{y_i}(1-g_i)^{1-y_i}p(\delta;\gamma_r)d\delta
\end{align}
where $g_i=g(x_i^t\beta+\delta_i)$, and $g$ is the link function mapping between the mixed linear score to the success rate. Usually, the optimization of $\mathcal{L}$ is done in an iterative manner and requires a lot of computational resources. However, in the specific case of LMM, where $\mathcal{F}$ is the Gaussian distribution and $g$ is the identity function, the integral in (\ref{Likelihood1}) can be written in a closed form that describes the marginal likelihood of $\beta$, $\varphi$, and $\gamma_r$, through the parameters of the variance component, encoded in the covariance matrix $V\equiv V(\varphi,\gamma_r) = \mathbb{C}ov(Y_{-k}|X_{-k})$. This yields the following NLL objective and the set of equations to the maximum likelihood estimator (MLE) of $\beta$: 
\begin{align*}
   NLL(\beta;T_{-k})&\propto \frac{1}{2}(Y_{-k}-X_{-k}\beta)^tV^{-1}(Y_{tr}-X_{-k}\beta) \\ S(\beta) :=\frac{\partial }{\partial \beta} NLL(\beta) &=  X_{-k}^tV^{-1}(Y_{-k}-X_{-k}\beta)=0,
\end{align*}
where $V\equiv V(\hat{\varphi},\hat{\gamma_r})$, where $\hat{\varphi}$ and $\hat{\gamma_r}$ are the MLEs of $\varphi$ and $\gamma_r$ respectively. The solution $\hat{\beta}_{T_{-k}}$ of the above set of equations can be written in closed form as follows:
\[\hat{\beta}_{T_{-k}}=( X_{-k}^tV^{-1}X_{-k})^{-1}X_{-k}^tV^{-1}Y_{-k}.\]

The closed form of $\hat{\beta}_{T_{-k}}$ in the linear model can be used to accelerate the calculation of $\hat{y}(x_{i};T_{b,-i})$ (or $\hat{y}(x_{i};T_{b_1,b_2,-i})$) within the bootstrap loop. We are motivated to extend this idea to GLMM in an approximated manner. When it comes to prediction, we first consider the scenario in which the predicted value $\hat{y}(x_{i};T_{-i})$ is a well-defined function of $x_{i}^t\hat{\beta}_{T_{-i}}$ (depending on the model performance criterion), such as in the spatial correlation scenario. Consequently,  the statistic $l_{b,i}$ in the bootstrap procedure is a well-defined function of $\hat{\beta}_{T_{b,-i}}$, where $\hat{\beta}_{T_{b,-i}}$ is the fitted estimator of $\beta$ based on the bootstrap training sample $T_{b,-i}$. Thus, our goal is to find an efficient but informative method to fit the estimator $\hat{\beta}_{T_{b,-i}}$ inside the bootstrap loop.


In order to fit $\hat{\beta}_{T_{b,-i}}$ efficiently, we embrace the idea of Quasi-Likelihood (QL) of the marginal likelihood of $\beta$ (discussed in \cite{wolfinger1993generalized} and \cite{ju2020laplace}), while using the parameter $\hat{\gamma}_r$ that was fitted once at the beginning of the algorithm using the whole data set $T$. We assume that $\hat{\beta}_{T_{b,-i}}$ obtained by optimizing $\mathcal{L}$, approximately satisfies the following QL equations:
\[S(\beta;T_{b,-i})= \left[X^tDV^{-1}\right]_{-i}(Y_{b,-i}-\mu_{-i})=0,\]
where the subscript $-i$ stands for the elimination of rows and columns in the same fold of observation $i$. The matrix $V\equiv V(\beta,\hat{\gamma}_r) = \mathbb{C}ov(Y|X;\beta,\hat{\gamma}_r)$, is the covariance matrix of $Y$, while conditioning on $X$, and assuming that the parameters of the model are $\beta$ and $\hat{\gamma}_r$. In particular, for any pair of indices $(i_1,i_2)$ such that $i_1\neq i_2$, we can write: $V_{i_1i_2} = \mathbb{C}ov_{\delta}\left(g(x_{i_1}^t\beta+\delta_{i_1}),g(x_{i_2}^t\beta+\delta_{i_2}) \right)$, and the diagonal elements also contain the iterative term: $\mathbb{E}_{\delta}\left[\mathbb{V}ar(y_i|x_i,\delta_i;\beta) \right]$. The vector $\mu \in \mathbb{R}^{n}$ satisfies $\mu_i=\mathbb{E}_\delta\left[g(x_i^t\beta+\delta_i)\right]$ and $D\in \mathbb{R}^{n \times n}$ is a diagonal matrix with elements $ D_{ii}=\frac{\partial \mu_i}{\partial( x_i^t\beta)}= \mathbb{E}_{\delta}\left[g'(x_i^t\beta+\delta_i)\right]$. 

Intuitively, the likelihood equations presented above aim to minimize the discrepancy between any label $y$ and its expectation $\mu$, while taking into account the correlation structure between observations, and averaging it over the distribution of the random effect $\delta$. The standard assumption of the QL approach is that the solution $\Breve{\beta}_{T_{b,-i}}$ of the QL equations is a good approximation of $\hat{\beta}_{T_{b,-i}}$ in terms of its correlation with $Y_{b,i}$. Nevertheless, finding an exact closed-form expression of $\Breve{\beta}_{T_{b,-i}}$ is not possible, and we use Taylor expansion around the vector $\hat{\beta}$ (evaluated globally only once base on the whole dataset $T$) to achieve the following quadratic approximation:
\begin{align} \nonumber
      \Breve{\beta}_{T_{b,-i}} &\approx\hat{\beta} +\left[(X^tDV^{-1}DX)^{-1} \right]_{-i}S(\hat{\beta};T_{b,-i}) = \tilde{\beta}_{b,-i},
\end{align}
where the entries of $V$, $\mu$, and $D$ are evaluated globally with $\hat{\beta}$, and by sampling random draws of $
\delta$ from the distribution $P_{\delta;\hat{\gamma}_r}$. 

The approximate solution $ \tilde{\beta}_{b,-i}$ can be calculated quickly for any draw of $Y_b$, and $\tilde{l}_{b,i}=L_{2}(\hat{y}(x_i^t\tilde{\beta}_{b,-i}))$ can be obtained immediately. The sample's covariance between $\tilde{l}_{i}$ and $y_{i}$ is denoted by $\widetilde{C}_i$, and the fast-approximated estimator of $w_{cv}$ is denoted by $\tilde{w}_{cv}$, and can be written as follows:
\[\tilde{w}_{cv} =\frac{1}{n}\sum_{i=1}^{n} \widetilde{C}_i =  \frac{1}{n}\sum_{i=1}^{n} \widehat{\mathbb{C}ov}\left( \tilde{l}_{i} ,y_{i} \right).\]
The approximated bias-corrected CV estimate is denoted by:
\[\widetilde{CV}_c=CV+\tilde{w}_{cv}.\] Interestingly, when the loss function is proportional to the NLL, and the model follows a canonical link, such as in linear and logistic regression, we have $l_{b,i}= L_{2}(\hat{y}(x_{i};T_{b,-i})) = x_i^t\hat{\beta}_{T{b,-i}}$, and the fast approximate statistic is $\tilde{l}_{b,i}=x_i^t\tilde{\beta}_{b,-i}$. In this particular scenario, and we can use the law of total covariance, and the standard properties of the covariance operator, to get the following simplified expression to $\widetilde{C}_i$ as follows:
\begin{align*}
    \widetilde{C}_i &= \widehat{\mathbb{C}ov}_{\delta_b}\left(
\mathbb{E}[x_i^t\tilde{\beta}_{b,-i}|X_{-i},\delta_b;\hat{\beta}], \mathbb{E}[y_{b,i}|x_i , \delta_b;\hat{\beta}] \right) \\ & = x_i^t\left[(X^tDV^{-1}DX)^{-1} X^tDV^{-1} \right]_{-i}\widehat{\mathbb{C}ov}_{\delta_b}\left(
\mathbb{E}[Y_{b,-i}|X_{-i},\delta_b;\hat{\beta}], g(x_i^t \hat{\beta} + \delta_{b,i})  \right) \\ &= 
x_i^t\left[(X^tDV^{-1}DX)^{-1} X^tDV^{-1} \right]_{-i}\widehat{\mathbb{C}ov}_{\delta_b}\left(
g\left(X_{-i} \hat{\beta} +\delta_{b,-i}\right), g(x_i^t \hat{\beta} + \delta_{b,i})  \right).
\end{align*}
We can see that in this case, we can avoid Step 3.b in the bootstrap procedure, when a vector $Y_b$ is drawn from $P_{Y|X , \delta_b;\hat{\beta}, \hat{\varphi}}$. Instead, we use the conditional mean of $Y$, based on the assumptions on the model and the estimates of the parameters. By this, we can reduce the variance of $\widetilde{C}_i$, which arises due to randomness in the process, and get a reliable estimation of the quantity of interest, with a smaller number of bootstrap iterations, $B$.

In some cases, the predicted value $\hat{y}(x_{i};T_{-i})$ may depend on some predictor of the specific realization of a latent random effect. Specifically, in the scenario of multiple-level clustered structure, assuming that observation $i$ is from entity $j_1$, the predicted value $\hat{y}(x_{i};T_{-i})$ is a well-defined function of $x_{i}^t\hat{\beta}_{T_{-i}}+ r_{te}^t \hat{u}_{T_{-i},j_1}$, where $\hat{u}_{T_{-1}}\in \mathbb{R}^{q_1}$ is the predictor of $u_{tr}$, based on the training sample $T_{-i}$. In this scenario the subscript $b$ denotes $b=(b_1,b_2)$. The estimation of $\hat{\beta}_{T_{b,-i}}$ stays consistent with the previously covered scenario, ensuring that the approximate estimator $ \tilde{\beta}_{b,-i}$ is still applicable. Consequently, the task now is to develop a fast approximation approach for estimating $\hat{u}_{T_{b,-i}}$.

Usually, the predictor $\hat{\eta}^t=(\hat{u}^t,\hat{s}^t)$ is obtained by maximizing the joint likelihood of $Y_{-i}$ and $\eta$, substituting $\beta$ with $\hat{\beta}_{T_{-i}}$, which can be written as follows:
\[\mathcal{L}(\eta,Y_{-i}|X_{-i};\hat{\beta}_{T_{-i}},\hat{\gamma}_r)\propto P_{Y_{-i}|X_{-i},\eta;\hat{\beta}_{T_{-i}}}\cdot P_{\eta;\hat{\gamma}_r},\]
where $\hat{\gamma}_r=(\hat{\sigma}^2_u,\hat{\sigma}^2_s)$. In the Quasi-Likelihood approach, this yields the following set of equations:
\[S(\eta)= \left[Q^tD_{\eta}V_{\eta}^{-1}\right]_{-i}(Y_{b,-i}-\mu_{\eta,-i})+V_{\hat{\gamma}_r}^{-1}\eta=0,\]
where $Q=\partial \delta / \partial \eta \in \mathbb{R}^{n \times (q_1+q_2)}$, and the matrix $V_{\eta}\equiv V(\hat{\beta}_{T_{-i}},\eta) = \mathbb{C}ov(Y|X,\eta;\hat{\beta}_{T_{-i}})$ is the covariance matrix of $Y$, while conditioning on $X$ and $\eta$. Moreover, $\mu_{\eta} = g\left(X\hat{\beta}_{T_{-i}}+\delta(\eta) \right)$, and $ D_{ii}=g'(x_i^t\hat{\beta}_{T_{-i}}+\delta_i(\eta))$. The matrix $V_{\hat{\gamma}_r}=diag((\hat{\sigma}_u^2 1_{q_1}^t , \hat{\sigma}_s^2 1_{q_2}^t ) )$ is the covariance matrix of $\eta$.

Using  a Taylor expansion around the generated vector $\eta_b^t=(u_{b_1}^t,s_{b_2}^t)$, and substituting $\hat{\beta}_{T_{-i}}$ with $\tilde{\beta}_{b,-i}$, we achieve the following approximation of the solution $\hat{\eta}_{T_{b,-i}}$ in the bootstrap procedure: 
\begin{align*}
    \hat{\eta}_{T_{b,-i}} &\approx \eta_{b} + \left( \frac{\partial S}{\partial \eta}|_{\eta=\eta_b}\right)^{-1} S(\eta_0):= \tilde{\eta}_{b,-i}, 
\end{align*}
where $\partial S /\partial \eta = \left[Q^tD_{\eta}V_{\eta}^{-1}D_{\eta}Q\right]_{-i}+V_{\hat{\gamma}_r}^{-1}$. Now, the approximate formula for the predictor $\hat{u}_{T_{b,-i}}$ is the first $q_1$ rows of $\tilde{\eta}_{b,-i}$: $\tilde{u}_{b,-i}=\left[\tilde{\eta}_{b,-i}\right]_{(1:q_1)}$. With the suggested formulas of $\tilde{\beta}_{b,-i}$ and $\tilde{u}_{b,-i}$, we can quickly calculate the estimator $\tilde{w}_{cv}$ of $w_{cv}$ in this scenario. The bias-corrected CV estimator obtained from the fast-approximated procedure is $\widetilde{\text{CV}}_c=\text{CV}+\tilde{w}_{cv}$

\subsection{Adaptations to Deep Learning}
As discussed in the introduction of this chapter (Section~\ref{C4_Intro} , some recent studies \citep{simchoni2021using,simchoni2023integrating,tran2020bayesian,xiong2019mixed} have suggested integrating the idea of a mixed-effects model into deep neural networks (DNNs), where the fixed-effects function $f_{\gamma_c}$ can be a non-trivial neural network with a set of parameters $\gamma_c$. The main idea (and also the main challenge) is to optimize the likelihood function $\mathcal{L}$ described in (\ref{Likelihood1}) with respect to $\gamma_c$ (instead of $\beta$) and $(\gamma_r,\varphi)$, as follows:
\[ (\hat{\gamma}_c,\hat{\gamma}_r,\hat{\varphi}) =   \underset{\gamma_c,\gamma_r,\varphi}{argmin} \left\{ \mathcal{L} \right\}, \]
where:
\begin{align}\label{DeepLikelihood1}
    \mathcal{L} &= \int p(Y_{-i}|X_{-i}, \delta ;\gamma_c,\varphi)p(\delta;\gamma_r)d\delta = \int \prod_{i=(k-1)\tilde{n}+1}^{k\tilde{n}} p(y_{i};x_{i},\gamma_c,\varphi|\delta)p(\delta;\gamma_r)d\delta.
\end{align}
In the specific case of binary classification,  we can write:
\begin{align}\label{DeepLikelihood2}
     (\hat{\gamma}_c,\hat{\gamma}_r) =   \underset{\gamma_c,\gamma_r}{argmin} \left\{ \mathcal{L} \right\} = \underset{\gamma_c,\gamma_r}{argmin} \left\{  \int \prod g_i^{y_i}(1-g_i)^{1-y_i}p(\delta;\gamma_r)d\delta \right\},
\end{align}
where $g_i=g(f_{\gamma_c}(x_i)+\delta_i)$.

In the context of this study, we consider the learning algorithm to be a black box, and we implement a cross-validation procedure when handling correlated data. In this case, the general methodology suggested for the estimation of $\hat{w}_{cv}$ outlined in this section can be applied, regardless of the complexity of $f_{\gamma_c}$. However, the computational resources required might be unbearable when $f_{\gamma_c}$ is a complex DNNs trained on a large-scale dataset. We propose two methods to address this problem and estimate the bias of the CV estimator in complex deep mixed models:
\begin{enumerate}
    \item {\bf  Hot-start training.} For each random training sample $T_{b,-i}$, start the training process with the parameters $\hat{\gamma}_c$, and $\hat{\gamma}_r$, which were already globally fitted once, based on the whole data $T$. Moreover, limit the number of iterations as necessary to control the required computational resources.

    \item {\bf Last-layer analysis.} Many DNN models have a fully connected layer with the appropriate activation function for the type of model as the output layer of $f_{\gamma_c}$. This layer can be viewed as a GLMM, while the rest of the network is fixed. Therefore, we can apply the fast approximate estimation outlined in Section \ref{Fast-Approximated} to the last layer, while all other parameters are treated as given. To be more precise, we write $\gamma_c=\{W,\beta\}$ where $W$ is the set of parameters of the hidden layers of $f_{\gamma_c}$ and $\beta$ is the parameter vector of the last layer. The output of the model in the bootstrap loop can be approximated as follows:
    \[\hat{y}(x_i;T_{b,-i}) \approx g\left(  \left(\tilde{f}(x_{i};\widehat{W})\right)^t   \tilde{\beta}_{b,-i} \right),\]
    where $\tilde{f}$ is the model up to the last layer, $\widehat{W}$ is the estimator of $W$ based on the whole data $T$, and $\tilde{\beta}_{b,-i}$ is calculated as described in Section \ref{Fast-Approximated} while substituting $X$ with $\tilde{f}(X;\widehat{W})$.

\end{enumerate}

Intuitively, we expect that both methods described above will result in an underestimation of the real bias term $w_{cv}$, due to the fact that changes of $\hat{y}(x_i;T_{b,-i})$ in $Y_{b,-i}$ are limited, compared to the case of a complete learning process from the beginning. However, constraining the variability of $\hat{y}(x_i;T_{b,-i})$ can guarantee that the adjusted estimator $\widetilde{\text{CV}}_c=\text{CV}+\tilde{w}_{cv}$ does not experience a higher variance relative to $\text{CV}$, while simultaneously diminishing some bias with respect to the true generalization error. The empirical study on real data in Section~\ref{Empirical} suggests that the use of $\widetilde{\text{CV}}_c$ is beneficial as an estimator of the real generalization error in terms of bias and variance.

\section{Empirical study}\label{Empirical}
We illustrate how the proposed methodology can be applied across four distinct learning scenarios:
\begin{enumerate}
    \item A simulated mixed logistic regression with two clustered correlation structure.
    \item A simulated mixed Poisson regression with two clustered correlation structure.
    \item A Convolutional Neural Network (CNN) model for smile detection, trained on celebrity images, with a clustering structure based on the celebrity's identity.
    \item A fully connected DNN model for the classification of asset properties, trained on Airbnb rental data, with a spatial correlation structure.
\end{enumerate}
In all scenarios involving classification, we consider three different criteria for the prediction error: cross-entropy loss, zero-one loss, and AUC measure. For each criterion, we compare the standard CV estimator with our suggested estimator $\widetilde{\text{CV}}_c$. In the following subsections, we present each scenario along with the primary outcomes.

\subsection{Mixed Logistic Regression}\label{Empirical_Res_Ber}
We perform an experiment on simulated data in which the outcome $y$ follows the Bernoulli distribution with two-level random intercept. Formally:
\begin{enumerate}
    \item Observation $y_{j_1,j_2,l}$ from the clusters $(j_1,j_2)$ satisfies $y_{j_1,j_2,l}\sim Ber\left(\frac{ exp\{LP_{j_1,j_2,l}\} }{1+exp\{LP_{j_1,j_2,l}\}} \right)$ for $j_1\in \{1,\cdots, 10 \}$, $j_2\in \{ 1,\cdots , 5\}$, and $l\in \{1,2, 3 \}$, where:
    \[LP_{j_1,j_2,l} = x_{j_1,j_2,l}^t\beta+u_{j_1}+s_{j_2},\]
    with $\beta=0.5\cdot \textbf{1}_p$, and $p=10$.
    
    \item The variance components satisfy $u_{j_1}\sim N(0, \sigma_u^2 )$, $s_{j_2}\sim N(0, \sigma_s^2 )$, where $\sigma_u=1$ and $\sigma_s=0.5$.
    
    \item Each random training sample is of size $n=110$, and the grouping variables $j_1 \in \{1,\cdots , 10\}$ and $j_2 \in \{1,\cdots , 5\}$ are assigned evenly so that $11$ observations are assigned to each group $j_1$ and $22$ observations are assigned to each group $j_2$, and $2$ or $3$ observations are assigned to each cluster $(j_1,j_2)$ at random. This data is stored in a one-hot encoded matrix $Z\in \{0,1\}^{n \times (q_1,q_2)}$, where $Z_{i,j_1}= Z_{i,10+j_2} =1$ and all other elements of the row $Z_i$ are zero if the $i$th observation is part of cluster $(j_1,j_2)$.

    \item The covariate matrix $X\in \mathbb{R}^{n \times p}$ is drawn from a multivariate Gaussian distribution with a general structure that takes into account the grouping relationships. 
    \item We draw latent variables: $u_{tr}\sim MN(0_{q_1},\sigma_u^2I_{q_1})$, $s_{tr}\sim MN(0_{q_2},\sigma_s^2I_{q_2})$ and they stored in $r_{tr}^t=(u_{tr}^t,s_{tr}^t)$.
    
    \item We draw the response vector $Y\sim Ber(g(X\beta+Zr_{tr}))$ where $g$ is the Sigmoid function.
\end{enumerate} 

The focus of this experiment is on the predictive capability of the relevant model from the GPBoost library \citep{JMLR:v23:20-322,GPBoostLib}, where the training sample size is $n=110$, and a 11-Folds-CV procedure aims to estimate the performance of the model with $100$ observations. The model is treated as a black box, producing an estimator $\hat{\beta}$ of $\beta$, estimates of $\sigma_s$ and $\sigma_u$, and predicted random effects $\hat{s}$ and $\hat{u}$ as output. For each $\Tilde{p}\in \{2,7,p\}$, we define the data set $T_{\tilde{p}}=(\mathcal{X}=X_{[0:\Tilde{p}]},Z,Y)$, where the subscript $[0:\Tilde{p}]$ refers to the first $\Tilde{p}$ columns of $X$, and $T_{\tilde{p}}$ is then divided into $11$ folds $\{T_k=(\mathcal{X}_k,Z_k,Y_k)\}_{k=1}^{11}$ to assess GenErr, CV, and $w_{cv}$ as detailed below:
\begin{itemize}
    \item The CV estimator is calculated by applying the same learning procedure (logistic regression) on each of the training folds ($T_{-k}$) and averaging over $L(y_i,\hat{y}({\mathcal{X}}_i;\hat{\beta}_{T_{-i}}))$, $i=1,\cdots,n$.
    \item The GenErr is evaluated by averaging over $L(y_i^{te},\hat{y}(x_i;\hat{\beta}_{T_{-i}}))$, where $y_i^{te}$ is the the $i$th index of $Y_{te}^t=(Y_{te,1}^t,\cdots , Y_{te,11}^t)$, such that $Y_{te,k}^t\sim  Ber(g(X_k\beta+Z_k r_{te,k}))$, $k\in \{1,\cdots, 11\}$, where each $r_{te,k}$ is a new realization of the latent random associated with the fold k, independent of $r_{tr}$ and the other $r_{te}$'s.
    \item To compute the estimator $w_{cv}$, we initially determine the matrix $V$ globally, and $(\mu,D)$ for each fold. This process involves generating random vectors $\widetilde{\mu}=X\hat{\beta}+Z\Tilde{r}$ and $\widetilde{\mu}_{\tilde{p}}=\mathcal{X}\hat{\beta}_{T_{-i}}+Z\Tilde{r}$ multiple times where $\Tilde{r}^t=(\Tilde{u}^t,\tilde{s}^t)$, $\Tilde{u}\sim MN(0_{q_1},\hat{\sigma}_u^2I_{q_1})$, and $s_{tr}\sim MN(0_{q_2},\hat{\sigma}_s^2I_{q_2})$, followed by computing the following empirical expectations:
    \[\mu = \widehat{\mathbb{E}}[g(\Tilde{\mu}_{\tilde{p}})] \hspace{2mm};\hspace{2mm} D =diag\left( \widehat{\mathbb{E}}[g'(\Tilde{\mu}_{\tilde{p}})] \right)  \hspace{2mm};\hspace{2mm}  V=\widehat{\mathbb{C}ov}\left( \tilde{\mu}\right) +diag\left( \widehat{\mathbb{E}}[g'(\Tilde{\mu})] \right).  \]
    We note that generally $\mathbb{V}ar(y|x,z,r)=g(x^t\beta+z^tr)(1-g(x^t\beta+z^tr))$, but when $g$ is the Sigmoid function, this expression simplifies to $\mathbb{V}ar(y|x,z,r)=g'(x^t\beta+z^tr)$.

    \item We then employ the bootstrap method, conducting $B=200$ iterations, using this formula for each iteration $b=1, \ldots, B$, and $k=\{1,\cdots, 11\}$:
    \[ \tilde{\beta}_{b,-k}= \hat{\beta} +\left[(\mathcal{X}^tDV^{-1}D\mathcal{X})^{-1} \right]_{-k} \left[\mathcal{X}^tDV^{-1}\right]_{-k}(Y_{b,-i}-\mu_{-k}). \]
\end{itemize}

We run the simulation on the Google Colab platform using a standard CPU. The running time for a single sample (including all three models and $B=200$ bootstrap iterations) is approximately $6$ seconds. 

\begin{figure}[ht!] 
\centering
\includegraphics[width=15cm]{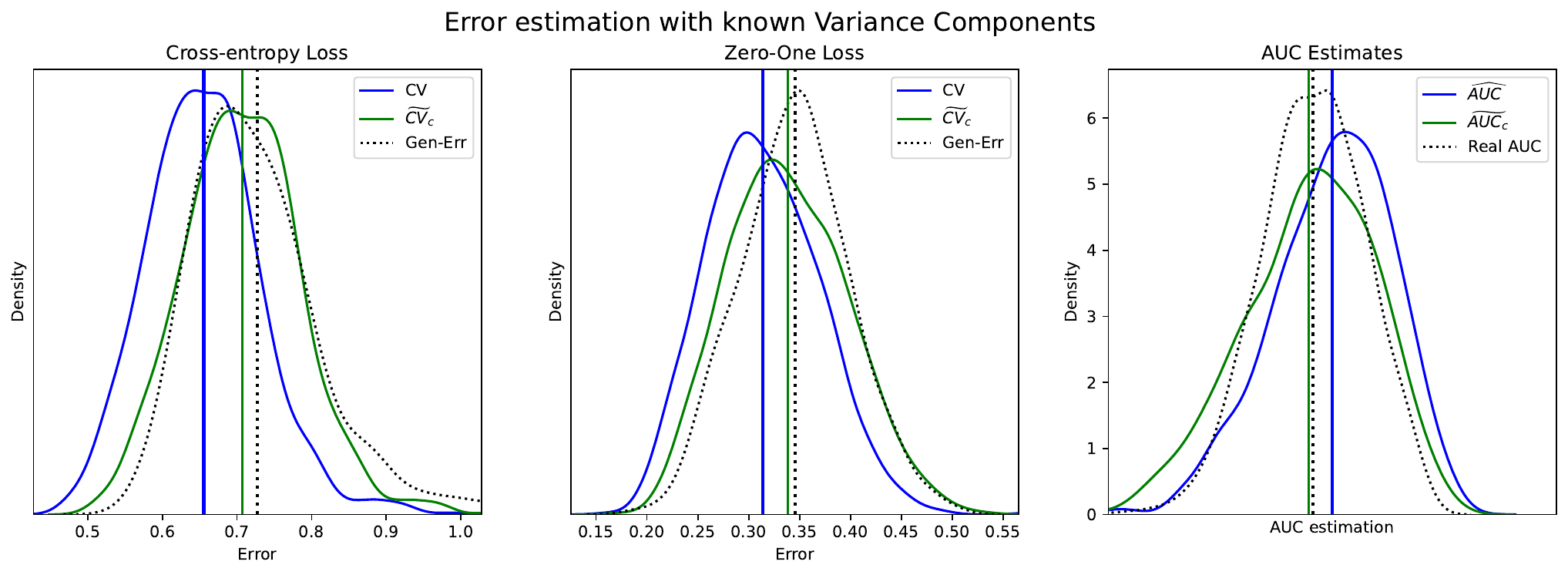}
\caption{Densities of $CV$, $\widetilde{CV}_c$ and the generalization error for different Loss functions in Logistic regression model, with known covariance parameters. The mean values are denoted by vertical lines.}\label{C4B1}
\end{figure}

\begin{figure}[ht!] 
\centering
\includegraphics[width=15cm]{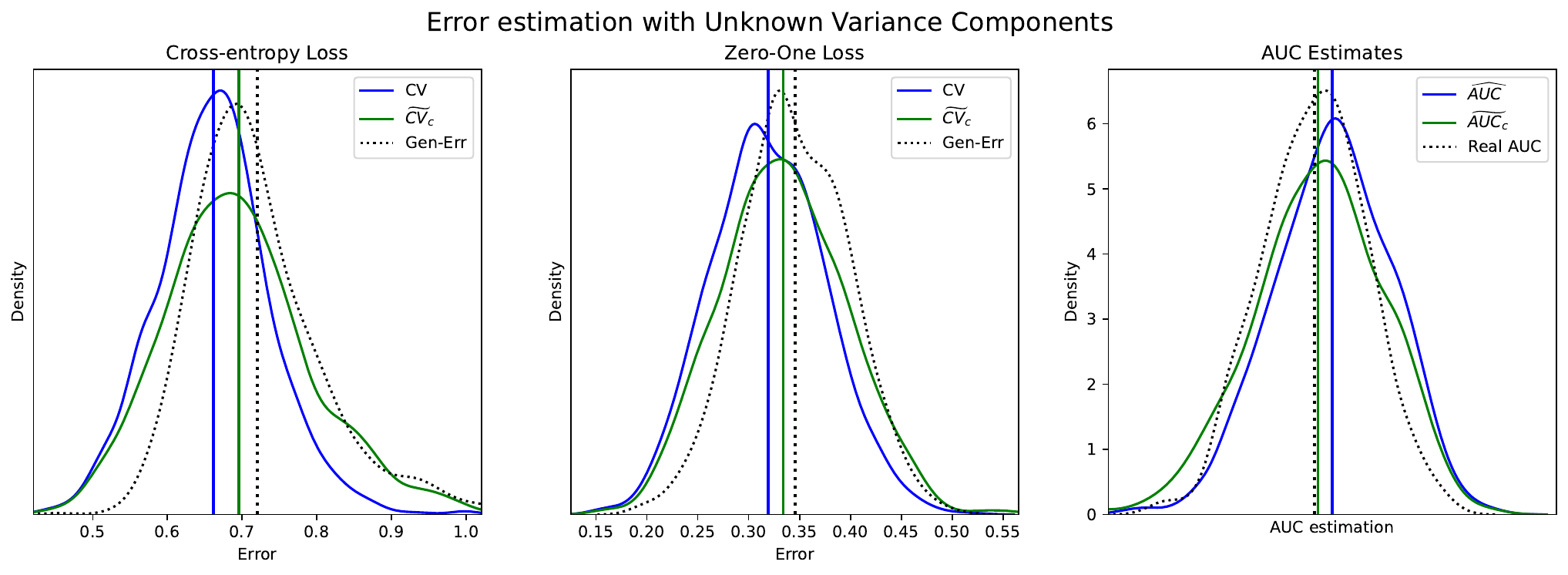}
\caption{Densities of $CV$, $\widetilde{CV}_c$ and the generalization error for different Loss functions in Logistic regression model, with unknown covariance parameters. The mean values are denoted by vertical lines.}\label{C4B11}
\end{figure}

In the described setting, we generate $2000$ training samples, and calculate the 11-Folds-CV estimator for each of them, as well as an estimation of the real Generalization Error. We consider the scenario in which the realizations of both $u_{j_1}$'s and $s_{j_2}$'s in the test set are not the same as in the training set. In Figure \ref{C4B1} we present the density of the statistics calculated on the $2000$ training samples in the specific case of $\tilde{p}=p$, with the covariance parameters $\sigma_u$ and $\sigma_s$ being known in the evaluation procedure of $\widetilde{CV}_c$. In Figure \ref{C4B11} we present the density of the statistics calculated with the covariance parameters $\sigma_u$ and $\sigma_s$ being unknown.

\begin{figure}[ht!] 
\centering
\includegraphics[width=12cm]{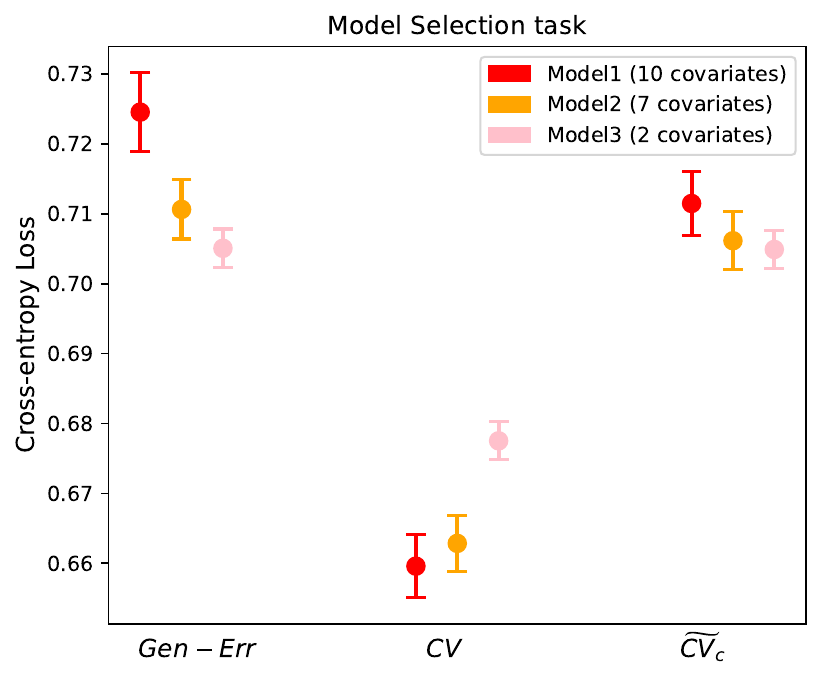}
\caption{Mean values of $CV$, $\widetilde{CV}_c$ and the generalization error and their 95\%  CIs for the three models.}\label{C4B2}
\end{figure}

We additionally examine the suitability of $\widetilde{\text{CV}}_c$ within the model selection framework, which considers three distinct models: the fully specified model incorporating all covariates in $x$, a model utilizing only the initial $7$ entries of $x$, and a model employing just the first $2$ entries. In figure \ref{C4B2}, it is evident that $\widetilde{\text{CV}}_c$ ranks the models correctly according to the true cross-entropy generalization error, whereas the conventional CV estimator ranks them in the reverse order.

\subsection{Mixed-Poisson regression}\label{Empirical_Res_Poiss}
We perform an experiment similar to the one in the previous section, but with the outcome $y$ follows the Poisson distribution, and with following parameters:
\begin{itemize}
    \item  $y_{j_1,j_2,l}\sim Poiss\left(exp\{0.2+x_{j_1,j_2,l}^t\beta+u_{j_1}+s_{j_2}\}\right)$,
    \item $j_1\in\{1,\cdots ,10\}$, $j_2\in\{1,\cdots ,5\}$ , $l\in\{1,2 ,3\}$, $\beta=0.2\cdot \textbf{1}_p$, $p=20$.
    \item $u_{j_1}\sim N(0, \sigma_u^2 )$, $s_{j_2}\sim N(0, \sigma_s^2 )$ where $\sigma_u=0.7$ and $\sigma_s=0.4$ are unknown.
    \item The covariates $x$ are generated as a Gaussian copula with uniform marginals such that $\mathbb{E}[x]=\textbf{0}_p$. 
    \item The loss function corresponds to the canonical NLL loss of a single draw: \[L(x_0,y_0;\hat{\beta})=exp\{x_0^T\hat{\beta}\}-x_0^T\hat{\beta} y_0.\]
\end{itemize}
The focus of this experiment is also on the predictive capacity of the relevant model from the GPBoost library \citep{JMLR:v23:20-322,GPBoostLib}. The model is treated as a black box, producing an estimator $\hat{\beta}$ of $\beta$, estimates of $\sigma_s$ and $\sigma_u$, and predicted random effects $\hat{s}$ and $\hat{u}$ as output.

In the described setting, we generate $K=1000$ training samples of size 110, and calculate the 11-Folds-$CV$ estimator for each of them, as well as an estimation of the Generalization Error of each fold, under two different scenarios.
\begin{enumerate}
    \item New random effects between the training set and the test set.
    \item The realizations of $\{u_{j_1}\}_{j_1=1}^{10}$ in the test set are the same as in the training set, but not the realizations of $\{s_{j2}\}_{j_2=1}^5$.
\end{enumerate}

We evaluated both the empirical estimator $\hat{w}_{cv}$ and the approximate estimator $\tilde{w}_{cv}$ of $w_{cv}$ for each sample and scenario, as described in Section \ref{MethodsForEst}. In the scenario with new random effects, $B=200$ bootstrap samples are used. In contrast, when the realizations of $\{u_{j_1}\}_{j_1=1}^{10}$ are identical in both the test and training sets, we utilize $B_1=20$ and $B_2=30$ bootstrap samples for the outer and inner loops, respectively. In both scenarios, the covariance parameters $\sigma_u$ and $\sigma_s$ are considered unknown during the evaluation of $\hat{w}_{cv}$ and $\tilde{w}_{cv}$.

\begin{figure}[ht!] 
\centering
\includegraphics[width=12cm]{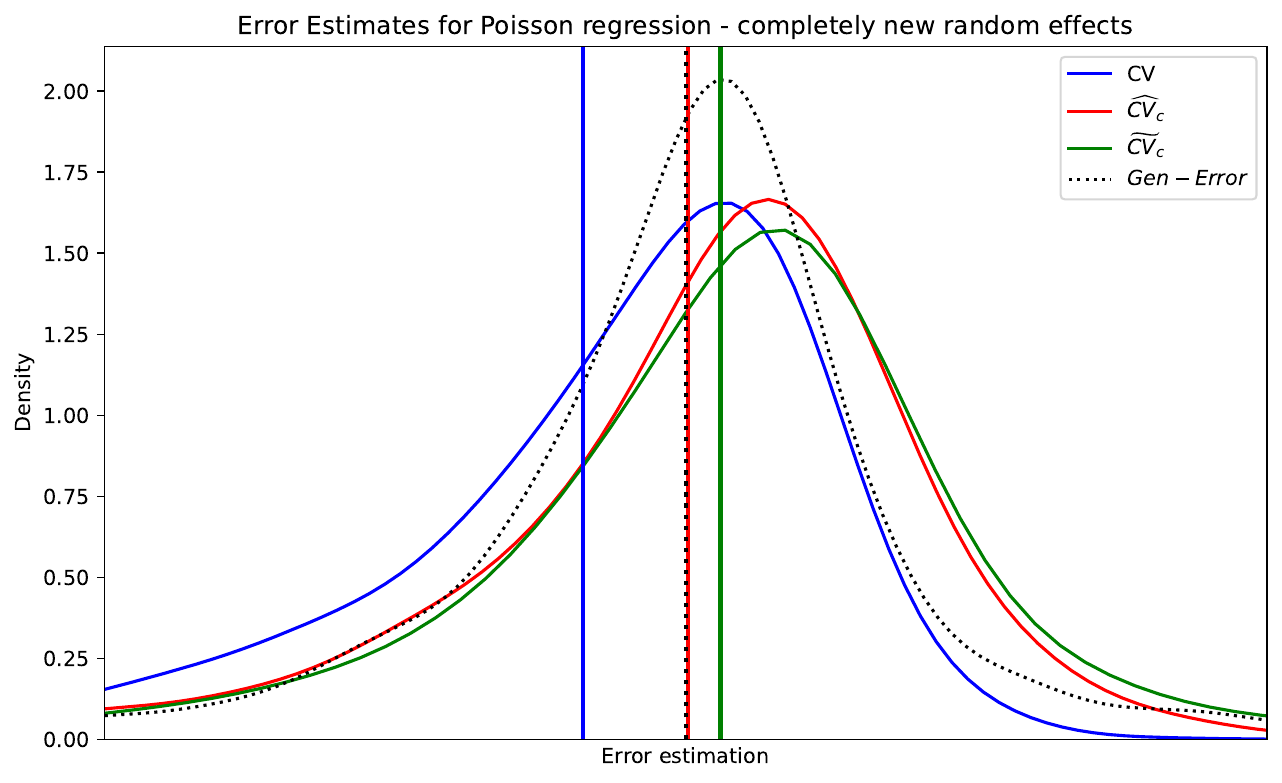}\caption{Densities of $CV$ , $\widehat{CV}_c$, $\widetilde{CV}_c$ and generalization error for Poisson regression model, in the scenario of completely new random effects. The mean values are denoted by vertical lines.}
\label{Poiss1}
\end{figure}

\begin{figure}[ht!] 
\centering
\includegraphics[width=12cm]{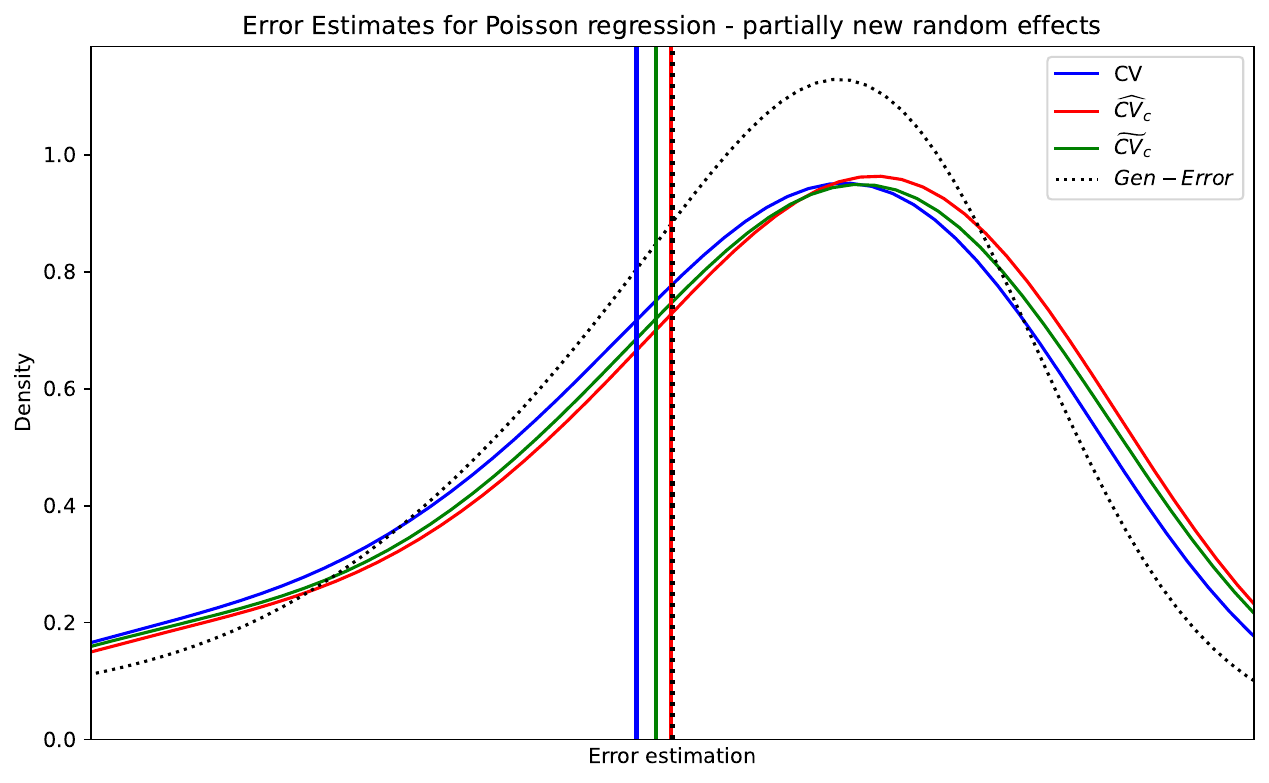}
\caption{Densities of $CV$ , $\widehat{CV}_c$, $\widetilde{CV}_c$ and generalization error for Poisson regression model, in the scenario of partially new random effects. The mean values are denoted by vertical lines.}
\label{Poiss2}
\end{figure}

Figures \ref{Poiss1} and \ref{Poiss2} show the empirical densities of $CV$, $\widehat{CV}_c=CV+\hat{w}_{cv}$, $\widetilde{CV}_c=CV+\tilde{w}_{cv}$ and the generalization error, along with their respective means indicated by vertical lines. It is evident that $\widehat{CV}_c$ is an unbiased estimator of the generalization error, while $CV$ is biased. Additionally, in the first scenario (Figure \ref{Poiss1}), the distribution of $\widehat{CV}_c$ has a smaller standard deviation (about $7\%$) and less skews to the left compared to $CV$, making its density more like the one of the generalization error. The $\widetilde{CV}_c$ estimator, however, appears to be biased in both scenarios, but much less than the standard $CV$ estimator. We attribute the bias of $\widetilde{CV}_c$ mainly to the discrepancy between the exponential loss function and its quadratic approximation.

\subsection{CNN model on the CelebA DataSet}\label{CelebAClassification}
We demonstrate the last-layer analysis method in CNN aimed at the task of smile detection in CelebA facial images. The CelebA dataset (\cite{liu2018large}) contains 202,599 facial images from 10,177 celebrities, each of which has between $1$ and $35$ images, with an average of $20$ images for a celebrity. Every image is labeled $1$ if the celebrity smiles and $0$ otherwise. We employ the same CNN architecture as in \cite{simchoni2021using}, which contains $4$ convolution layers followed by $2$ fully connected layers. We aim to evaluate how well the model can predict the outcomes for new random identities that it did not encounter during the training stage, where the training dataset comprises $q = 100$ identities. It is presumed that for each identity $j \in \{1, \ldots, q = 100\}$ from the training sample, there is an impact from a random effect $s_j$ distributed as $N(0, \sigma_s^2)$, and that the link function $g$ that specifies the model is the Sigmoid function. Under this setting, the contribution of identity $j$ to the negative log likelihood (NLL) can be written as follows:
\begin{align*}
    \mathbb{L}_j(\gamma_c,\sigma_s^2)= - log \left\{ \frac{1}{\sqrt{\pi}}\int exp\left[\sum_{l=1}^{n_j} y_{jl} \zeta_{jl}(u)-log\left(1-e^{\zeta_{jl}(u)}\right) 
 \right] e^{-u^2}du \right\}, 
\end{align*}
where $n_j$ is the number of observations of identity $j$, and $\zeta_{jl}(u)=f_{\gamma_c}(x_{jl})+\sqrt{2}\sigma_s u$. To approximately minimize $\mathbb{L}_j(\gamma_c,\sigma_s^2)$ with respect to $(\gamma_c,\sigma_s^2)$, \cite{simchoni2023integrating} proposed employing Gauss-Hermite quadrature. This approach results in the following objective function:
\begin{align*}
    \mathbb{L}_j(\gamma_c,\sigma_s^2) \approx - log \left\{  \sum_{k=1}^{K_{GH}} \frac{w_k}{\sqrt{\pi}} exp\left[  \sum_{l=1}^{n_j} y_{jl} \zeta_{jl}(\nu_k)-log(1-e^{\zeta_{jl}(\nu_k)}) 
 \right]\right\}, 
\end{align*}
where $\nu_k$ is the $k$th zero of the Hermite polynomial of degree $K_{GH}$, and $w_k$ is the associated weigh. With the stated objective function formula above, we are able to train the model using stochastic gradient descent (SGD), selecting batches based on identities. We provide an implementation of the described learning procedure in the PyTorch library. 

\begin{figure}[ht]
\centering
\includegraphics[width=16cm]{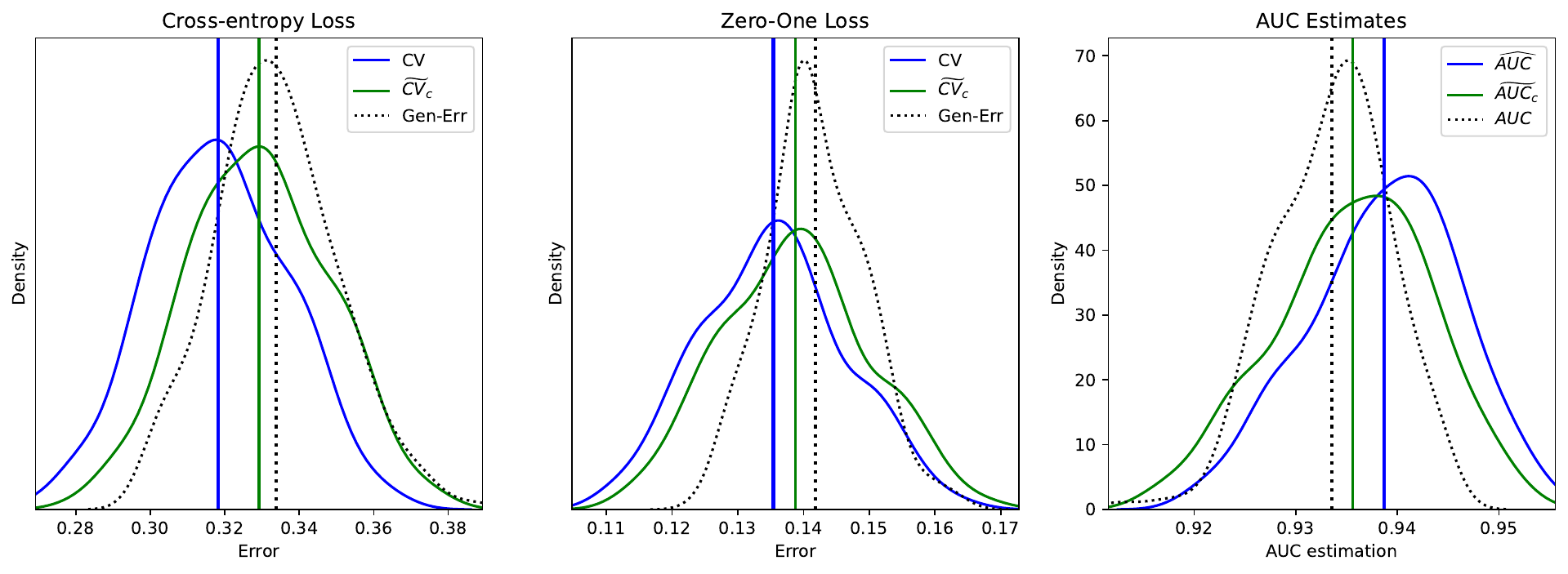}
\caption{Densities of $CV$ , $\widetilde{CV}_c$ and generalization in the CNN model for smile detection on CelebA dataset. The mean values are denoted by vertical lines.}\label{C4D1}
\end{figure}

The described approach is applied on the whole training dataset, which generally includes $q =100$ randomly chosen identities and about $2000$ images in total. We perform $80$ epochs, to fit the estimators $\hat{\gamma}_c$ and $\hat{\sigma}_s^2$. Initially, $\hat{\sigma}_s^2$ is set to $1$, with learning rates of $10^{-4}$ for $\hat{\sigma}_s^2$, and $10^{-5}$ for network $\gamma_c$. Subsequently, the estimator $\Tilde{w}_{cv}$ is evaluated using the last-layer analysis approach, similar to the procedure detailed in the Mixed Logistic Regression simulations. The CV estimator is evaluated using a standard $10$-Fold cross-validation procedure, and the GenErr is evaluated folds-wise on distinct test set of random $900$ identities. We run the simulation on the Google Colab platform using a Tesla T4 GPU. The duration of processing a single sample, taking into account the CV procedure, is approximately $50$ minutes. In Figure~\ref{C4D1} we present the results of $200$ random samples. We can see that the estimator $\Tilde{w}_{cv}$ effectively captures most of the bias observed between the $CV$ estimate and the true generalization error. 

\subsection{DNN model on AirBNB data set}
We apply the Hot-start technique to a fully connected DNN model utilizing a dataset consisting of 50,000 Airbnb properties. This data set was collected and preprocessed in a similar manner to \cite{rezazadeh2021airbnb}, based on the \cite{AirbnbDataset} public Airbnb dataset for New York City, with specific modifications, resulting in the creation of the "\textit{AirBNBlistings\_cleaned.csv}" dataset. The model is trained using properties situated in randomly chosen small regions, each spanning $100$ meters, with labels assigned as $1$ if the property features hot water. Spatial correlation is incorporated using a Gaussian kernel based on property distances, with its parameters co-optimized with the model's. The sampling and evaluation mechanism is similar to Section \ref{CelebAClassification}, aiming to assess the model's ability to predict labels for new properties in different regions. The empirical marginal success rate of the outcome variable, "hot water", is $0.43$. Each training set $T$ comprises $100$ randomly selected geographical "blocks" each measuring $100$ meters, containing on average approximately $2000$ listings. Any test set, on the other hand, consists of $500$ randomly chosen geographical blocks.

A fully connected network, $f_{\gamma_c}$, serves as the learning model, featuring two hidden layers with ReLU activations, containing $100$ and $20$ neurons, respectively. It is presumed that a listing $i$ of the training set, featuring hot water with probability $g(f_{\gamma_c}(x_i)+\delta_{T,i})$ where $g$ is the Sigmoid function and $\delta_T\in \mathbb{R}^n$ is the vector of random effects associated with the data. Moreover, we assume that $K_{\delta_T,i_1,i_2}=\mathcal{K}(z_{i_1},z_{i_2};\sigma_s^2;\tau^2)=\sigma_s^2 exp\{-\tau^2||z_{i_1}-z_{i_2} ||_2^2\}$, where $K_{\delta_T}\in \mathbb{R}^{n\times n}$ is the covariance matrix of $\delta_T$. Under this setting, the negative log likelihood (NLL) can be written as follows:
\begin{align*}
    \mathbb{L}(\gamma_c,\sigma_s^2,\tau^2)= - log \left\{ \frac{1}{\pi^{n/2}}\int _{u \in \mathbb{R}^n}exp\left[\sum_{i=1}^{n} y_{i} \zeta_{i}(u)-log\left(1-e^{\zeta_{i}(u)}\right) 
 \right] e^{-u^tu}du \right\}, 
\end{align*}
where $\zeta_{i}(u)=f_{\gamma_c}(x_{i})+\sqrt{2}\left[K_{\delta_T}^{0.5}u\right]_i$. To achieve an approximate minimization of $\mathbb{L}$ with respect to $(\gamma_c,\sigma_s^2,\tau^2)$, we introduce a Gauss-Hermite quadrature approximation performed in two dimensions for pairs of observations. By employing this method, we derive the subsequent objective function related to the observations $i_1$ and $i_2$:
\begin{align*}
    \mathbb{L}_{i_1,i_2}(\gamma_c,\sigma_s^2,\tau^2) \approx - log \left\{  \sum_{k_1,k_2} \frac{w_{k_1}w_{k_1}}{\pi} exp\left[  \sum_{l=1}^{2} y_{i_{l}} \zeta_{i_{l}}(\nu_{k})-log(1-e^{\zeta_{i_l}(\nu_{k})}) 
 \right]\right\}, 
\end{align*}
where $\nu_k^t=(\nu_{k_1},\nu_{k_2})$ is the combinations of $k_1$th and $k_2$th zeros of the Hermite polynomial, and $(w_{k_1},w_{k_1})$ are the associated weighs. Moreover, $\zeta_{i_l}(\nu_{k})=f_{\gamma_c}(x_{i_i})+\sqrt{2}\left[K_{i_1,i_2}^{0.5}\nu_k\right]_{l}$, where:
\[K_{i_1,i_2}=\begin{pmatrix}
\sigma_s^2 & \sigma_s^2 e^{-\tau^2||z_{i_1}-z_{i_2} ||_2^2} \\
\sigma_s^2 e^{-\tau^2||z_{i_1}-z_{i_2} ||_2^2
} & \sigma_s^2 
\end{pmatrix}.\]	
Utilizing the given objective function formula, we can employ SGD to optimize the model with respect to the parameters $(\gamma_c,\sigma_s^2,\tau^2)$, by randomly selecting batches of size $2$. To enhance the efficiency of the procedure, we adopted the approach outlined in \cite{jackel2005note}, considering only the combinations $(k_1,k_2)$ for which $w_{k_1}w_{k_1}$ exceeds a specified threshold value. To maintain the stability of SGD, we first perform an initial optimization of the network parameters $\gamma_c$ using the typical cross-entropy loss function, before fully incorporating the entire NLL objective function. We provide an implementation of the described learning procedure in the PyTorch library.

The described approach to learning is applied to the entire training data set for $200$ epochs with the typical cross-entropy loss function (with batches of size 100), and $10$ epochs of pairwise full-parameter optimization. Initially, $\hat{\sigma}_s^2$ and $\tau^2$ are set to $1$ (we note that the distance unit is scaled to 1 nautical mile), with learning rates of $10^{-5}$ for $(\hat{\sigma}_s^2,\tau^2)$, and $10^{-4}$ to fit the parameters of the network $\gamma_c$. Evaluation of the estimator $\Tilde{w}_{cv}$ is performed using the Hot-Start analysis technique, employing $100$ standard epochs across $B=20$ bootstrap iterations. We run the simulation on the Google Colab platform utilizing a standard CPU. The duration of processing a single sample, taking into account the CV procedure, is approximately $35$ minutes.

\begin{figure}[ht]
\vspace{.15in}
\centerline{\includegraphics[width=13cm]{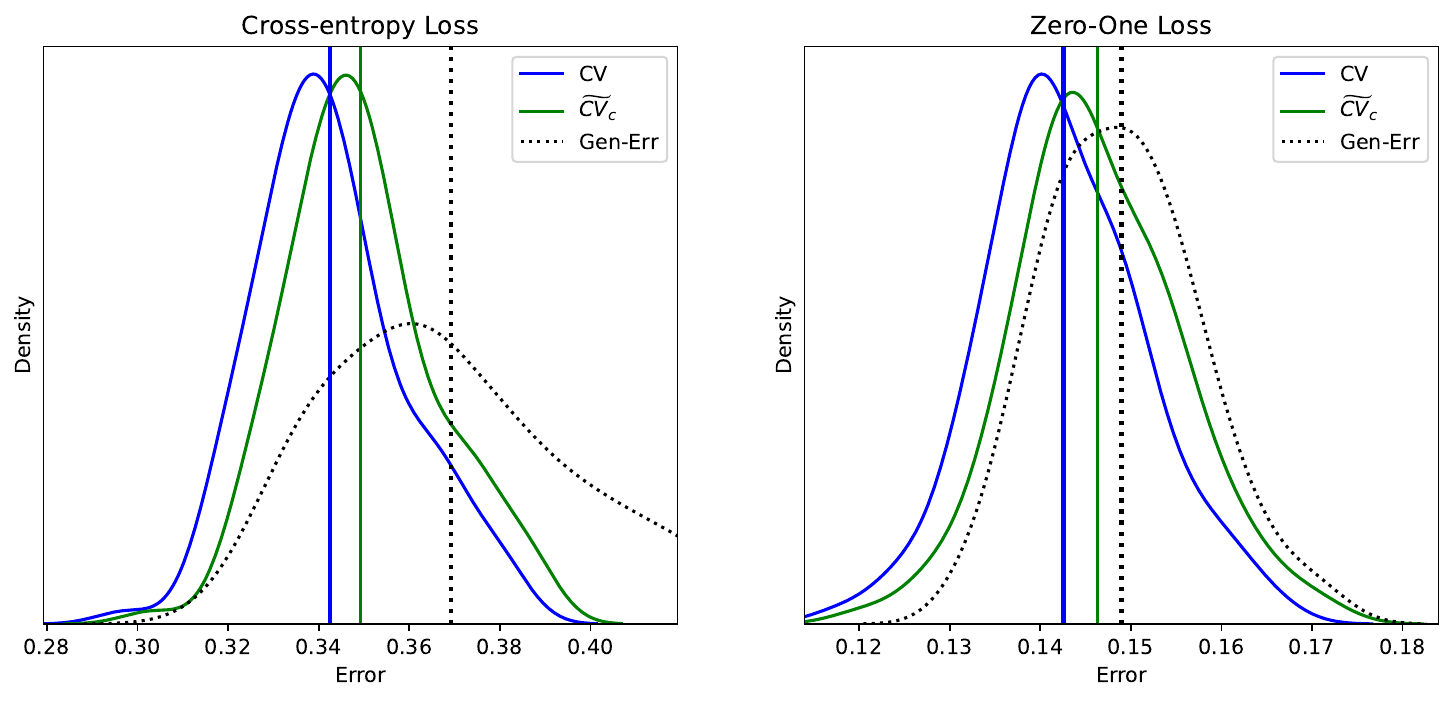}}
\vspace{.15in}
\caption{Densities of $CV$ , $\widetilde{CV}_c$ and generalization in the CNN model for hot water feature detection on AirBNB dataset. The mean values are denoted by vertical lines.}\label{D2}
\end{figure}

In the results, Figure \ref{D2}, we can see that the estimator $\widetilde{CV}_c$ is effective in reducing the bias of the CV estimator, especially in the case of zero-one loss. In the case of cross-entropy loss, GenErr is very nosy with a long tail to the right, which is not well captured by the approximate estimator $\tilde{w}_{cv}$.

\section{Conclusion}
In this chapter, we have illustrated the way that cross-validation (CV) methods can be tailored to accommodate correlation structures that exist within datasets. This is particularly relevant in scenarios where the distribution of random effects differs between the datasets used for training, employed in CV, and those used for predictions. Our techniques involve a careful examination of how covariance structures impact model assessment, leading to a precise formulation of the bias between standard CV approaches and the generalization error. Moreover, our approach has been shown to be effective both in simulation studies and when applied to real datasets, in conjunction with cutting-edge learning algorithms.

The primary finding related to the formulation of $w_{cv}$, as detailed in Equation \ref{WcvForm2}, holds true under the specific condition where $x_{te}$ and $x_1$ are governed by the identical distribution $P_x$. In contrast, the earlier formulation outlined in Equation \ref{WcvForm1} is more comprehensive and accommodates scenarios where $P_{x_{te}|X_{-1}}$ is not necessarily equal to $P_{x_1|X_{-1}}$. Future research could tackle the challenge of estimating $w_{cv}$ in more general and realistic contexts involving varied correlation structures, such as covariate shifts or other intricate relationships.

\bibliography{Bib}

\end{document}